\documentclass[twocolumn,prc,floatfix,showpacs,nopreprintnumbers,nofootinbib,%
superscriptaddress]{revtex4}
\usepackage{mathrsfs}
\usepackage{amssymb}
\usepackage{amsmath}
\usepackage{graphicx}
\usepackage[breaklinks]{hyperref}
\usepackage{mciteplus}
\usepackage[latin1]{inputenc}

\newcommand{\xt}{{\mathbf{x}_T}}

\newcommand{\bt}{{\mathbf{b}_T}}

\newcommand{\bta}{{\mathbf{b}_{T1}}}
\newcommand{\btap}{{\mathbf{b}'_{T1}}}
\newcommand{\btb}{{\mathbf{b}_{T2}}}

\newcommand{\xtp}{{\mathbf{x}'_T}}
\newcommand{\btp}{{\mathbf{b}'_T}}

\newcommand{\ztp}{{\mathbf{z}'_T}}
\newcommand{\yt}{{\mathbf{y}_T}}
\newcommand{\zt}{{\mathbf{z}_T}}
\newcommand{\ut}{{\mathbf{u}_T}}
\newcommand{\vt}{{\mathbf{v}_T}}
\newcommand{\utp}{{\mathbf{u}'_T}}

\newcommand{\pti}{{\mathbf{p}_{T1}}}
\newcommand{\ptii}{{\mathbf{p}_{T2}}}
\newcommand{\qt}{{\mathbf{q}_T}}
\newcommand{\kt}{{\mathbf{k}_T}}
\newcommand{\kti}{{\mathbf{k}_{Ti}}}
\newcommand{\ktj}{{\mathbf{k}_{Tj}}}

\newcommand{\ptrig}{{p_{T}^\textrm{trig}}}
\newcommand{\pass}{{p_{T}^\textrm{ass}}}

\newcommand{\nt}{{\mathbf{n}_T}}

\newcommand{\ptt}{{p_T}} 
\newcommand{\ktt}{{k_T}} 
\newcommand{\ntt}{{n_T}} 
\newcommand{\utt}{{u_T}} 

\newcommand{\Deltat}{{\boldsymbol{\Delta}_T}}

\newcommand{\ud}{\, \mathrm{d}}

\newcommand{\Tr}{\, \mathrm{Tr} \, }

\newcommand{\nc}{{N_\mathrm{c}}}

\newcommand{\cf}{C_\mathrm{F}}

\newcommand{\nr}[1]{(\ref{#1})}

\newcommand{\gev}{\ \textrm{GeV}}

\newcommand{\mb}{\ \textrm{mb}}

\newcommand{\qs}{Q_\mathrm{s}}
\newcommand{\qso}{Q_\mathrm{s0}}
\newcommand{\qsa}{Q_\mathrm{sA}}
\newcommand{\qsp}{Q_{\mathrm{s}p}}
\newcommand{\lqcd}{\Lambda_{\mathrm{QCD}}}
\newcommand{\as}{\alpha_{\mathrm{s}}}

\newcommand{\fig}{Fig.~}

\newcommand{\eq}{Eq.~}
\newcommand{\se}{Sec.~}
\newcommand{\eqs}{Eqs.~}
\newcommand{\re}{Ref.~}
\newcommand{\res}{Refs.~}

\newcommand{\seff}{{ \sigma_\textrm{eff} }}

\begin{document}

\author{T. Lappi}
\affiliation{
Department of Physics, %
 P.O. Box 35, 40014 University of Jyv\"askyl\"a, Finland
}

\affiliation{
Helsinki Institute of Physics, P.O. Box 64, 00014 University of Helsinki,
Finland
}
\author{H. M\"antysaari}
\affiliation{
Department of Physics, %
 P.O. Box 35, 40014 University of Jyv\"askyl\"a, Finland
}

\title{
Forward dihadron correlations in deuteron-gold collisions
with a Gaussian approximation of JIMWLK
}


\begin{abstract}
We compute dihadron correlations in forward deuteron-gold or 
proton-gold collisions. The running coupling BK equation is used to calculate
the energy dependence of the dipole cross sections and
extended to higher-point Wilson line correlators using a factorized
Gaussian approximation.
Unlike some earlier  works we include both
 the ``inelastic'' and ``elastic'' contributions to the dihadron 
cross section. We show that 
the double parton scattering contribution is included in our calculation
and obtain both an away side peak 
that roughly agrees with experimental observations and an estimate for the
azimuthal angle-independent pedestal.
We find that nonlinear effects for momenta close to the saturation scale
are clearly visible in the away side peak structure
\end{abstract}

\maketitle

\section{Introduction}

At high energy or, equivalently small~$x$, the interactions
of hadrons are expected to be dominated by  nonlinear strong 
color fields. A convenient effective theory approach to studying
these color fields is provided by the Color Glass Condensate
(for reviews see e.g.~\cite{Gelis:2010nm,Lappi:2010ek}).
This effective theory is based on the division
of the QCD degrees of freedom into a static (in light cone time)
large~$x$ part, which is described as a color source, and 
the small~$x$ gluons as a color field radiated by these sources.
The rapidity scale separating the two is, of course, arbitrary,
and should be chosen to correspond to the rapidity scale
probed in the particular physical process being looked at. 
From the requirement that physical cross sections
must not depend on this arbitrary scale one can derive
renormalization group equations describing the rapidity 
dependence of the probability distribution of the 
color sources. The general, leading $\ln x$, equation 
for the probability distribution of different color
charge configurations  is known by the acronym 
JIMWLK (see \re\cite{Mueller:2001uk} and references therein). 
The mean field approximation of JIMWLK, known as the
Balitsky-Kovchegov~\cite{Balitsky:1995ub,Kovchegov:1999yj,Kovchegov:1999ua}
equation, is widely used in phenomenological applications.
The BK equation describes the rapidity dependence of the unintegrated 
gluon distribution  and can loosely
be thought of as describing the rapidity dependence of the 
mean number of gluons instead of the whole probability 
distribution encoded in the JIMWLK equation.

The BK equation is usually enough when studying single inclusive
cross sections in a dilute-dense collision such as DIS
or proton-nucleus collisions at forward rapidities. Understanding
the initial state of a heavy-ion collision, on the other hand, necessitates
the, eventually numerical, solution of a much more complicated
interacting system~\cite{Krasnitz:1998ns,Lappi:2011ju}. Therefore
these very inclusive quantities do not provide direct experimental access
to all of the physics described by the JIMWLK equation
(for a discussion of potential observables in diffractive DIS that 
go beyond the mean-field BK approximation see e.g.
\re\cite{Marquet:2010cf}). It has recently  become more
evident that multiparticle correlations could provide precisely
such an observable.

Quite generically  multiparton 
correlations are more sensitive to the detailed 
dynamics of the colliding objects than single particle distributions. 
An attractive observable, 
measurable at both RHIC and the LHC, are dihadron correlations
in the forward rapidity region in proton-nucleus or deuteron-nucleus
 collisions. In this kinematical regime
the dilute and relatively well understood large~$x$ part of the proton 
provides with a calibrated probe of the nonlinear small~$x$ gluon fields
in the target nucleus. By varying the rapidity one can study different
values of~$x$ and the transverse momentum dependence of the correlation
should be directly sensitive to the dominant intrinsic transverse
momentum scale in the target, the saturation scale $\qs$.

Indeed measurements of correlations between two forward 
dihadrons  measured in dAu collisions at 
RHIC~\cite{Braidot:2011zj,Adare:2011sc}
seem to show indications of  ``initial state'' 
or ``cold nuclear matter'' effects that are significantly stronger than
in pp collisions or at central 
rapidities~\cite{Adler:2006hi,Citron:2009eu,Braidot:2009ji}.
The upcoming LHC proton-lead collisions will 
provide more opportunities to study these phenomena in a wider
kinematical range.
These observations have provided an impetus for renewed interest
in the gluonic correlations included in the JIMWLK
evolution~\cite{Dumitru:2010mv,Dumitru:2010ak,%
Dumitru:2011vk,Iancu:2011ns,Iancu:2011nj,Kovner:2011pe,Kovner:2010xk}.
In particular it was argued~\cite{Dumitru:2011vk,Iancu:2011ns,Iancu:2011nj}
that the result of a full JIMWLK evolution, also at finite $\nc$,
can quite accurately  be captured by the so called Gaussian 
approximation, relating higher-point Wilson line correlators
to the two-point function. 
Thus in the Gaussian approximation one can construct observable 
cross sections using the solution of the BK equation alone, which
is much more convenient in practice than solving the full JIMWLK equation.
These recent theoretical developments were not fully reflected
in the pioneering calculations of dihadron correlations 
in~\cite{Marquet:2007vb,Albacete:2010pg} 
(see also the more recent work~\cite{Dominguez:2011wm,Stasto:2011ru}
where a $\ktt$-factorized approximation is derived
in a certain kinematical limit). 
The main purpose of this paper is
to implement the Gaussian approximation, which so far has only been tested 
for particular coordinate space configurations, in a full calculation 
of the dihadron correlation

We shall first, in \se\ref{sec:sinc}, 
discuss the description of the single inclusive
hadron spectrum as a baseline that should be consistently 
described by the same calculation and used to constrain the initial conditions
of evolution in rapidity. 
We then, in \se\ref{sec:dihad}, 
recall the expression of the dihadron cross section derived in
\re\cite{Marquet:2007vb} using  the 
light cone perturbation theory formalism.
In \se\ref{sec:dps} we show that the dihadron cross section
as derived in \re\cite{Marquet:2007vb} contains a logarithmically
infrared divergent part, which has been overlooked in the previous literature.
We identify this as a double parton scattering
contribution describing the independent scattering of two 
partons already present in the dilute projectile.
The double parton scattering contribution 
must be consistently subtracted from
the correlated cross section and absorbed into a separate contribution
that depends on additional nonperturbative information about the 
projectile in the form of a double parton distribution.
We then discuss in \se\ref{sec:multipt} 
the expressions for the Wilson line correlators used in this
work. In \se\ref{sec:res} we present our results for the 
dihadron cross section. We show that using the full expression
for the dihadron cross section
enhances the away-side peak by a large factor compared to
approximations used in the previous literature. The qualitative 
features of the results in the previous literature remain,
most prominently a strong dependence of the correlation on 
$\ptt/\qs$, which leads to a large difference between deuteron-gold 
and proton-proton collisions.
Technical details on the DPS limit are relegated
to Appendix~\ref{app:dps} and on the impact parameter dependence
to Appendix~\ref{app:b}.

\section{Baseline description of the target:   
BK evolution and single inclusive spectra}
\label{sec:sinc}
\begin{figure}[tb]
\begin{center}
\includegraphics[width=0.5\textwidth]{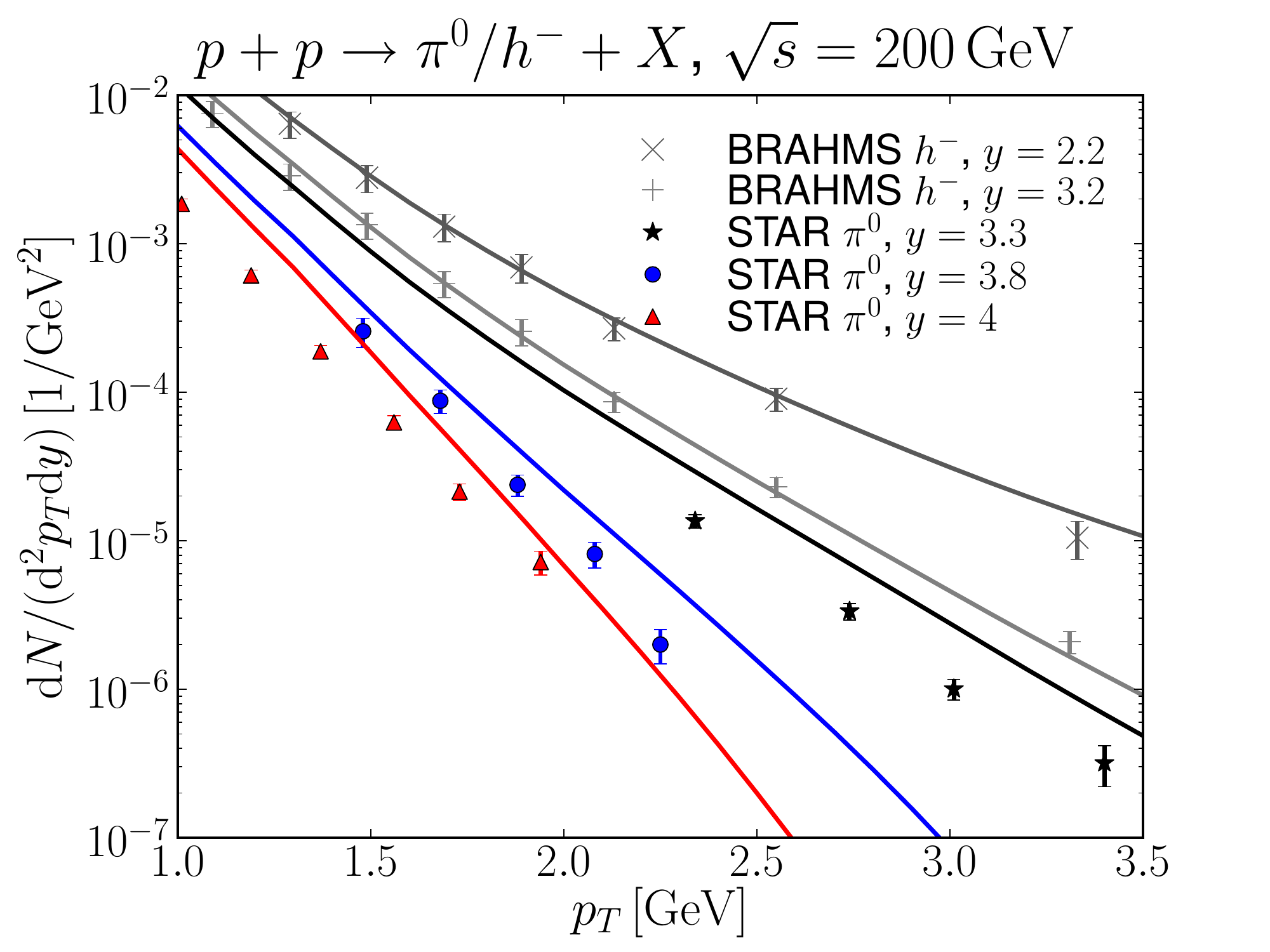}
\end{center}
\caption{
The identified hadron spectra at forward rapidity compared to the
experimental data from proton-proton collisions from the 
STAR~\cite{Adams:2006uz} 
and BRAHMS~\cite{Arsene:2004ux}  collaborations. 
} \label{fig:ppsinc}
\end{figure}

\begin{figure}[tb]
\begin{center}
\includegraphics[width=0.5\textwidth]{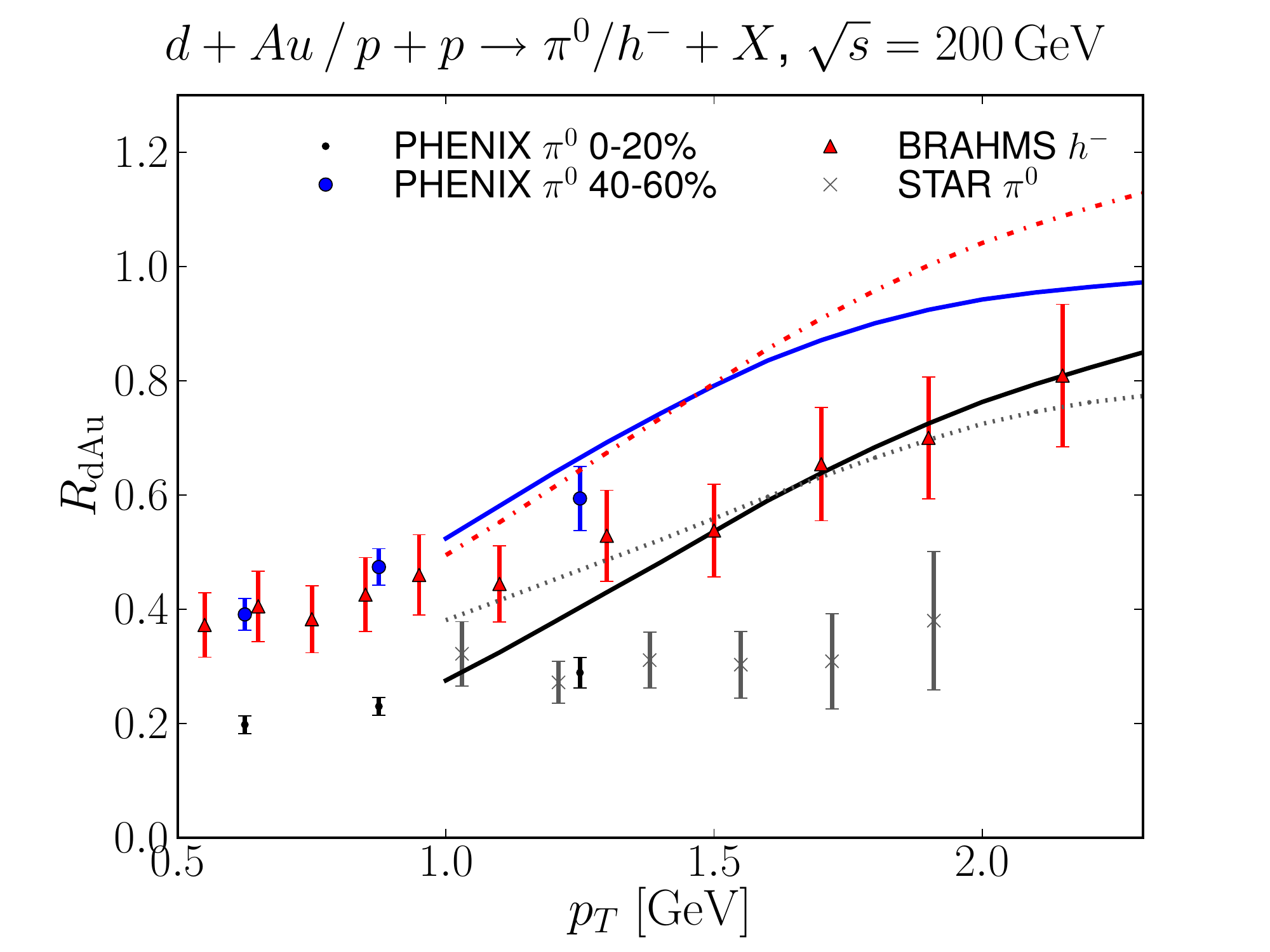}
\end{center}
\caption{
The nuclear modification factor $R_\textrm{dAu}$
using our MV dipole cross section parametrization. 
The experimental results shown are
PHENIX~\cite{Adare:2011sc}  centrality dependent (solid curves, $3<y<3.8$)
and BRAHMS~\cite{Arsene:2004ux} ($y=3.2$, upper dashed line) 
and STAR~\cite{Adams:2006uz} ($y=4$, lower dashed line)
minimum bias results.
The calculations are the spectra obtained using
\eq\nr{eq:sinc}. 
} \label{fig:rdau}
\end{figure}

 We shall here use the dipole cross sections obtained from  solving numerically the 
BK evolution equation using the Balitsky running coupling 
prescription~\cite{Balitsky:2006wa}
(see the comparison of different running coupling schemes
 in \re\cite{Albacete:2007yr}).
 The evolution equations must be 
supplemented with an initial condition at a starting rapidity (or $x$).
Ideally one would like to fully constrain the initial conditions of BK/JIMWLK
evolution by a comparison to small~$x$ DIS data. Due to the large amount of precise 
data available from HERA, the expectation value of the dipole cross section in a 
proton is quite well constrained. Several ways to extend these parametrization to 
nuclei exist in the literature, and in practice it is useful to also compare
with single inclusive hadron production data in pA and dAu to further
constrain the nuclear dipole cross sections. We shall here first recall
the (``hybrid formalism'') equations used to calculate the single inclusive
hadron yields at forward rapidity. We then discuss the 
MV model parametrization of the initial condition  
used in this paper  and compare it to the experimental 
forward single inclusive spectra.

The single inclusive yields for quark and gluon production are given 
by~\cite{Dumitru:2001jn,Dumitru:2002qt,Dumitru:2005gt} 
\begin{eqnarray}\label{eq:dsigmasq}
\frac{\ud N^{hA\to qX}}{\ud y  \ud^2 \qt    } 
& =&  \frac{1}{(2\pi)^2} xq(x)  S(\qt),
\\ \label{eq:dsigmasg}
\frac{\ud N^{hA\to g X}}{\ud y  \ud^2 \qt    } 
& =&  \frac{1}{(2\pi)^2} xg(x)  \widetilde{S}(\qt),
\end{eqnarray}
where $S(\qt)$ is the Fourier-transform of the 
fundamental representation dipole operator
$\langle \hat{D}\rangle$ and $\widetilde{S}(\qt)$ the corresponding
quantity in the adjoint representation. Here 
$xq(x)$ and $xg(x)$ are standard collinear parton distribution
functions describing the large~$x$ partons from the dilute projectile, 
for which we use the CTEQ NLO~\cite{Lai:2010vv}  parametrization. 
To get the single hadron spectrum we  convolute the parton-level cross section
with the DSS~\cite{deFlorian:2007aj}  fragmentation function, resulting in 
\begin{equation}\label{eq:sinc}
\frac{\ud N^{hA\to h' X}}{\ud y  \ud^2 \qt    } 
= \sum_{i}\int \frac{\ud z}{z^2}  
\frac{\ud N^{hA\to i X}}{\ud y  \ud^2 (\qt/z)    } 
D_{i\to h'}(z,\mu^2).
\end{equation}
Throughout this work the scale $\mu$ at which the parton distribution function, fragmentation function and strong coupling constant $\as$ are evaluated is chosen to be the transverse momentum of the produced hadron in single inclusive production and transverse momentum of the leading hadron in dihadron production.

As an initial condition we have used the MV model
parameters used in \re\cite{Albacete:2010bs} to fit single inclusive
hadron production data at RHIC. The initial dipole cross section is 
\begin{equation} \label{eq:bkinitc}
S(r)_{x=x_0} = \exp\left\{
 -\frac{r^2\qso^2}{4}  \ln \left( e + \frac{1}{r^2\lqcd^2}\right)\right\},
\end{equation}
 with an initial saturation scale
$\qso^2=0.2\gev^2$ at $x_0=0.007$ for the proton, as in 
\re\cite{Albacete:2010bs}.
A comparison of the resulting yields in proton-proton collisions
to the forward results of the STAR~\cite{Adams:2006uz} and BRAHMS~\cite{Arsene:2004ux} 
experiments is shown in \fig\ref{fig:ppsinc}.
While the description of the data is not perfect, we are
confident that this is a realistic enough baseline for understanding 
nuclear modifications to the dihadron yield.
Note that there are no arbitrary normalization 
 $K$-factors used in this work.

 Estimates based on nuclear geometry would 
suggest~\cite{Kowalski:2007rw,Tribedy:2011aa}
that the initial saturation scale $\qso^2$ should be even as large as
$A^{1/3}\sim 6$ times that of the proton (for minimum bias 
collisions). Based on both experimental data and the general expectation
that saturation effects should vanish for $\ptt \gg \qs$, it
would also seem natural for the  
nuclear modification factor $R_\textrm{dAu}$ to approach $1$ for
high $\ptt$ at least at midrapidity.
Assuming that the transverse area of a proton also at 
small $x$ is identified with the total inelastic nucleon-nucleon 
cross section, this would happen with 
$\qsa^2 \sim N_{\textrm{bin}}^{\textrm{pA}} \qsp^2 \approx 3.6 \qsp^2$
for a minumum bias dAu collision\footnote{
One would generically expect $\qs^2$ to be proportional to the number of 
overlapping nucleons, a quantity which is in practice encoded 
as the number of binary collisions in a Glauber model for a proton-nucleus
collision. In this paper we will assume 
$N_{\textrm{bin}}^{\textrm{pA}}=N_{\textrm{bin}}^{\textrm{dA}}/2$ for simplicity.
Note that $N_{\textrm{bin}}^{\textrm{pA}}$  is also the number of participant 
nucleons in the target nucleus in a pA collision.}.

 However, fits to identified hadron spectra,
especially at forward rapidities tend to favor
a smaller nuclear saturation scale. In the calculation 
of \re\cite{Albacete:2010bs}, for example, a calculation of 
hadron spectra in deuteron-gold collisions is made with 
an initial nuclear saturation scale $\qsa^2 \approx 2.3 \qsp^2 $ 
(comparing $\qsa$ and $\qsp$ at $x=0.007$, taking into account the
different $x_0$ used for protons and for nuclei). These 
saturation scales give a good description of the BRAHMS charged
hadron spectra at $\eta=2.2$ and $\eta=3.2$. 

For the most forward $\pi^0$ data the nuclear modification ratio 
$R_\textrm{dAu}$ seems to decrease faster
with $y$ in the data than predicted by running coupling BK evolution. 
For example, in the calculation 
of \re\cite{Albacete:2010bs} the STAR very forward
$\eta=4$ data requires an additional suppression by $K$-factors
$0.4$ for pp and $0.3$ for dAu and a similar effect is
seen in \cite{Fujii:2011fh,Tribedy:2011aa}. This is problematic for the 
dihadron correlation, for which the STAR forward data comes precisely at
this same kinematical region.

In view of these uncertainties,
we shall not attempt to provide the best possible fit to all the existing 
single inclusive particle production data in this work.
 Instead we use as an initial condition for nuclei 
the same parametrization \nr{eq:bkinitc} as for protons,
with a nuclear saturation scale
obtained by straightforward 
$N_{\textrm{bin}}$ scaling. This is essentially equivalent 
to the Monte Carlo rcBK model of \re\cite{Albacete:2010ad},
where the saturation scale $\qs^2$ is taken to be proportional
to the number of overlapping nucleons in a position in the
transverse plane. In particular it leads, by construction,
to a nuclear modification factor $R_\textrm{dAu}$ approaching 
unity at high transverse momenta, when the unintegrated
parton distribution is probed at the scale of the initial 
condition at $x_0$. For minimum bias collisions
this leads to an initial nuclear saturation scale
$\qso^2=0.72\gev^2$, and for a 0-20\% centrality 
class $\qso^2=1.51\gev^2$, using the 
$N_{\textrm{bin}}$ values from \re\cite{Adare:2011sc}.

We show in Fig.~\ref{fig:rdau} the resulting 
forward  $\pi^0$ nuclear modification ratio 
$R_\textrm{dAu}$  compared to the
PHENIX~\cite{Adare:2011sc}  centrality dependent 
and BRAHMS~\cite{Arsene:2004ux} and 
 STAR~\cite{Adams:2006uz} minimum bias data. 
Let us now point the reader to some features in this plot. 
Firstly, none of the nuclear data compared to
here has been used as an input in the calculation; instead the
initial condition for BK evolution has been set from a simple and 
straightforward $N_{\textrm{bin}}$ scaling of the saturation scale.
Secondly, the STAR minimum bias data exhibits approximately
as much suppression as the most central PHENIX data. Although the two
are at slightly different rapidities, it seems quite generically
impossible to simultaneously describe both data sets in 
a unique parametrization as simple as ours. A more detailed treatment of
minimum bias collisions by averaging over different centralities
in the calculation (instead of a single $\qs$ value characterising
minimum bias collisions) could also be expected to improve this description.
Note that the negative hadron nuclear modification ratio 
becomes $>1$ at high $\ptt$ due to the difference between the deuteron
and proton probes (a relative excess of $d$ quarks in the deuteron 
over the proton leads to an enhancement in negative particle production 
independently of the target).
All in all, given that this is a pure prediction for the nuclear modifiction
of single particle production, without arbitrary $K$ factors or adjusting
the  parameters to the nuclear data, we consider this as an adequate
parametrization for the purposes of understanding the basic features
of dihadron correlations, which is the main purpose of this paper.

\section{Dihadron correlations} \label{sec:dihad}

We shall consider the scattering process for forward dihadron production as
a large~$x$ quark with momentum $p^+$ from the probe deuteron or proton, 
propagating eikonally through the target nucleus or proton. It can 
radiate a gluon with momentum $k^+=zp^+$ and is left with a 
longitudinal momentum $q^+ = (1-z)p^+$. 
In the high energy limit the scattering of both the quark 
and the gluon can be described by an eikonal approximation, where 
they pick up a phase given by a Wilson line in the color field of the
target.  The detailed derivation  of the double inclusive
cross section is performed in \re\cite{Marquet:2007vb} and 
results in the following expression for the $qA\rightarrow qgX$ cross section:
\begin{widetext}
\begin{equation}
\label{eq:dihadron-xs}
\begin{split}
\frac{\ud \sigma^{qA\to qgX}}{\ud k^+ \ud^2 \kt \ud q^+ \ud^2 \qt } 
&= \as \cf \delta(p^+-k^+-q^+) 
\int \frac{\ud^2 \xt}{(2\pi)^2} \frac{\ud^2 \xtp}{(2\pi)^2} 
\frac{\ud^2 \bt}{(2\pi)^2} \frac{\ud^2 \btp}{(2\pi)^2} 
e^{i \kt \cdot(\xtp - \xt)} e^{i\qt \cdot(\btp-\bt)} \\
&\quad\times \sum_{\alpha\beta \lambda} 
\phi_{\alpha\beta}^{\lambda*}(\xtp-\btp) \phi_{\alpha\beta}^\lambda(\xt- \bt) 
\{ S^{(4)}(\bt,\xt,\btp,\xtp) - S^{(3)}(\bt,\xt,\ztp) \\
&\quad - S^{(3)}(\zt,\xtp,\btp) 
+ S^{(2)}(\zt,\ztp) \} ,
\end{split}
\end{equation}
with $\zt = z \xt +(1-z) \bt$ and likewise, $\ztp = z \xtp + (1-z)\btp$. 
Here the target is described by expectation values of the Wilson line operators
\begin{align}
\label{eq:s4}
S^{(4)}(\bt,\btp,\xt,\xtp) &= \frac{2}{\nc^2-1}\left \langle 
\Tr \left(V(\bt) V^\dagger(\btp) t^d t^c \right) [U(\xt) U^\dagger(\xtp)]^{cd}\right \rangle 
\\
\label{eq:s3}
S^{(3)}(\bt,\xt,\ztp) &= \frac{2}{\nc^2 -1 } \left \langle
 \Tr \left( V^\dagger(\ztp) t^c V(\bt) t^d \right) U^{cd}(\xt) \right \rangle 
\\
\label{eq:s2}
	S^{(2)}(\zt,\ztp) &= \frac{1}{\nc} 
\left \langle \Tr \left( V(\zt)V^\dagger(\ztp)\right)\right \rangle.
\end{align}
\end{widetext}
The momenta of the produced gluon and quark are  $k$ and $q$ respectively.
Likewise, $\xt,\xtp$ should be interpreted as the transverse position
of the gluon, $\bt,\btp$ of the quark after the scattering and $\zt,\ztp$
of the quark before the scattering; in 
the amplitude and the complex conjugate respectively. 
The wave function $\phi$ describes $q\to qg$ splitting in coordinate space, 
and its expression in the massless limit is given in \eq \eqref{eq:wavef-prod}.
In our numerical calculations we set the quark mass $m_q=0.14\gev$, but 
the finite quark mass has little effect on the final results.

The scattering
amplitude is a sum of two terms corresponding to the radiation of
the gluon happening before and after the interaction with the target. 
Out of the three operators \nr{eq:s4}, \nr{eq:s3}, \nr{eq:s2} in $S^{(4)}$
the gluon is radiated before the interaction and in  $S^{(2)}$ after, with
$S^{(3)}$ corresponding to the interference between gluon radiation before 
and after the target.

The operators  \nr{eq:s4}, \nr{eq:s3}, \nr{eq:s2} are expressed in terms of
fundamental and adjoint representation Wilson lines, denoted here by $V$ and $U$,
 respectively.
In the following we denote the dipole and quadrupole operators by
\begin{equation}
\hat{D}(\xt-\yt) \equiv \frac{1}{\nc} \Tr \left( V(\xt) V^\dagger(\yt) \right)
\end{equation}
\begin{equation}
\hat{Q}(\xt,\yt,\ut,\vt) = {1\over \nc} \Tr \left( V(\xt) V^\dagger(\yt) 
V(\ut) V^\dagger(\vt) \right),
\label{eq:quad}
\end{equation}
and their expectation values as $S\equiv S^{(2)}\equiv \langle \hat{D}\rangle$
and $Q\equiv \langle \hat{Q}\rangle$. In terms of these 
the higher-point correlators  in \eq\nr{eq:dihadron-xs} can 
be written in the form quoted in \res\cite{Dumitru:2010mv,Dumitru:2011vk}
\begin{multline}
\label{eq:wlinecorr4}
S^{(4)}(\bt,\btp,\xt,\xtp) = 
\Bigg\langle - \frac{1}{\nc^2-1}\hat{D}(\bt,\btp)
\\ 
+ \frac{\nc^2}{\nc^2-1}
\hat{D}(\xt,\xtp) 
	   \hat{Q}(\bt,\btp,\xtp,\xt) 
\Bigg\rangle
\end{multline}
\begin{multline}
\label{eq:wlinecorr3}
S^{(3)}(\bt,\xt,\ztp) =
\\
\Bigg\langle  \frac{\nc^2}{\nc^2-1} \hat{D}(\bt,\xt)\hat{D}(\xt,\ztp) 
 - \frac{1}{\nc^2-1} \hat{D}(\bt,\ztp)
\Bigg\rangle
.
\end{multline}
We will discuss explicitly in \se\ref{sec:multipt} the Gaussian approximation
used in this paper to obtain the Wilson line correlators in 
\eqs\nr{eq:wlinecorr4} and~\nr{eq:wlinecorr3}.
Computing the dihadron correlation in LHC kinematics will require taking
into account also the gluon-initiated channel, because the probe
$x$ will not be as large. This would require calculating
an eight-point function of Wilson lines, which we leave for future work.

\section{Double parton scattering contribution}
\label{sec:dps}

In the massless limit the wave function product appearing in 
\eq\nr{eq:dihadron-xs} is given by
\begin{equation}
\label{eq:wavef-prod}
\sum_{\alpha\beta\lambda} \phi_{\alpha\beta}^{\lambda*}(\utp)
			  \phi_{\alpha\beta}^\lambda(\ut) = 
\frac{8\pi^2}{k^+} \frac{\ut \cdot \utp}{|\ut|^2|\utp|^2}(1+(1-z)^2),
\end{equation}
with $\ut = \xt-\bt$ and $\utp = \xtp-\btp$.

In the full dihadron cross section one must also include
a  ``double parton scattering'' (DPS) contribution that corresponds to the 
 independent scattering of two partons
from the probe. This contribution, whose importance for understanding
the $\Delta \varphi$-independent pedestal  of the double inclusive 
cross section, was  emphasized in~\cite{Strikman:2010bg}.
 We shall show  that \eq\nr{eq:dihadron-xs} contains
a part of this (DPS)   contribution, which must be subtracted from it
to avoid double counting. This is in contrast with 
\re\cite{Dominguez:2011wm,Stasto:2011ru} 
where, instead of the full cross section
\eq\nr{eq:dihadron-xs}, one considers only the ``correlation limit'' in which 
the DPS contribution vanishes. As discussed in \re\cite{Dominguez:2011wm}
 the full cross section can differ 
significantly from the correlation limit when the transverse momenta
of the produced particles are comparable to the saturation scale, which is
precisely the regime that is interesting for observing
the effects of large color fields in the target.

In the limit where the gluon is far away from the quark:
$|\bt - \btp|\sim |\xt - \xtp| \sim 1/\qs,$ 
$\utt \equiv |\bt-\xt|\gg 1/\qs$ the Wilson lines of the quark and the gluon
are uncorrelated. Therefore the expectation values in 
$S^{(4)}$ factorize into
a product of an adjoint representation dipole at the location of the
gluon and a 
fundamental representation one at the location of the quark. 
We call this kinematical regime the ``DPS'' limit since, as we will argue 
in the following, it corresponds to a quark and a gluon, present already
in the wavefunction of the incoming dilute projectile, scattering
 independently off the target. 
Factorizing the expectation values
and using the fact that the expectation values must be color singlets 
we can write $S^{(4)}$ in the DPS limit as
\begin{multline}
\label{eq:s4dps}
S^{(4)}(\bt,\xt,\btp,\xtp) 
\underset{\textrm{DPS}}{\approx}
S^{(4)}_\textrm{DPS}(\bt,\xt,\btp,\xtp) 
\\
\equiv \frac{2}{\nc^2-1} 
\left \langle \Tr \left(  V(\bt) V^\dagger(\btp) t^d t^c \right) \right \rangle
\left \langle[U(\xt) U^\dagger(\xtp)]^{cd} \right \rangle
\\
=
\frac{\nc^2}{\nc^2-1}
\left \langle \hat{D}(\bt,\btp)\right \rangle
\left \langle \hat{D}^2(\xt,\xtp) - \frac{1}{\nc^2}\right \rangle,
\end{multline}
where 
\begin{equation}
\widetilde{S}(\xt-\xtp)
\equiv
\frac{\nc^2}{\nc^2-1}
\left \langle 
\hat{D}^2(\xt,\xtp) - \frac{1}{\nc^2} \right\rangle
\end{equation}
 can be identified
as the two point function in the adjoint representation, normalized
to $\widetilde{S}(\xt=\xtp)=1$.
 The cross terms $S^{(3)}$ vanish in the DPS limit. This is easily seen by 
noting that $z$ is finite as it is fixed by the final state kinematics and thus
all three coordinates $\bt,\xt$ and~$\ztp$ (or $\btp,\xtp$ and~$\zt$) are much
more than $1/\qs$ apart. Consequently $S^{(3)}$ factorizes into a product of
three single Wilson line expectation values, which are all zero.
The Wilson line operator 
corresponding to  gluon emission after the target, $S^{(2)}$, is finite in the
DPS limit. It does not, as we show in Appendix~\ref{app:dps}, 
give a divergent contribution 
to the double inclusive cross section even for  massless quarks.

In the massless case  the
integral \nr{eq:dihadron-xs} diverges logarithmically in the DPS
limit, because $S^{(4)}$ is nonzero and the 
wave function product has a large power law tail at large $\ut,\utp$. 
This logarithmic divergence is shown explicitly in Appendix~\ref{app:dps}.
 Physically this means that the quark emits a very small 
transverse momentum gluon. The quark and gluon subsequently 
scatter independently off the target.  A simple lifetime argument shows that
the emission of the quark happens $\Delta x^+ \sim z(1-z) p^+ \utt^2$ before the 
interaction with the target. Thus the contribution from $\utt\to \infty $
 corresponds to a splitting happening a long time before the interaction 
with the target.
This logaritmically 
divergent contribution must  be regulated by 
confinement scale physics in the wavefunction of the projectile. It is in fact exactly 
the kind of contribution that is represented by double parton 
scattering~\cite{Strikman:2001gz,Blok:2010ge}. 
The confinement scale physics of the correlations  in the large~$x$ 
projectile 
is not calculable in our formalism. We shall instead add it as an additional 
explicit DPS contribution. To avoid double counting we must subtract the equivalent
contribution from our \eq\nr{eq:dihadron-xs}, which will render it manifestly IR finite.
We do this by replacing $S^{(4)}(\bt,\xt,\btp,\xtp)$ in \eq\nr{eq:dihadron-xs}
by
\begin{multline}
S^{(4)}_{\textrm{sub}}(\bt,\xt,\btp,\xtp) = S^{(4)}(\bt,\xt,\btp,\xtp) 
\\
- \theta\left(|\xt-\bt| - \frac{1}{\lqcd} \right)
\theta\left(|\xtp-\btp| - \frac{1}{\lqcd}\right)
\\
\times 
S^{(4)}_\textrm{DPS}(\bt,\xt,\btp,\xtp) 
\end{multline}
with $S^{(4)}_\textrm{DPS}$ given by \eq\nr{eq:s4dps}.

After convoluting with the valence quark distribution in the probe deuteron or proton 
this correlated, $\Delta \varphi$-dependent part of the double differential 
yield becomes
\begin{widetext}
\begin{multline}
\label{eq:dAuqg-xs}
\frac{\ud N^{h_a h_b \to qgX}_{\textrm{sub}}}{\ud y_q \ud^2 \kt \ud y_g \ud^2 \qt } 
= 
\as \cf x q(x,Q^2) (1-z) (1+(1-z)^2) 
\frac{1}{S_\perp}
\int \frac{\ud^2 \xt}{(2\pi)^2} \frac{\ud^2 \xtp}{(2\pi)^2} 
\frac{\ud^2 \bt}{(2\pi)^2} \frac{\ud^2 \btp}{(2\pi)^2} 
\\
e^{i \kt \cdot(\xtp - \xt)} e^{i\qt \cdot(\btp-\bt)} 
 8\pi^2 
\ 
\frac{(\xt-\bt)\cdot(\xtp-\btp)}{(\xt-\bt)^2(\xtp-\btp)^2}
\\
\{ S^{(4)}_{\textrm{sub}}(\bt,\xt,\btp,\xtp) - S^{(3)}(\bt,\xt,\ztp) 
 - S^{(3)}(\zt,\xtp,\btp) 
+ S^{(2)}(\zt,\ztp) \}.
\end{multline}
\end{widetext}
Here we have moved from a cross section to a differential yield by dividing 
with the transverse area of the target $S_\perp$.
Because the integrand is translationally invariant as a whole, 
one out of the four integrals over the transverse plane in 
\eq\nr{eq:dAuqg-xs} is performed trivially, canceling the 
factor $1/S_\perp$
 (see also the discussion on the impact parameter dependence in 
Appendix \ref{app:b}).
 We assume the Wilson line correlators to be independent
of the overall impact parameter; thus the dependence
on centrality only comes through the  Wilson line correlator,
which should correspond to an average one for the desired centrality class.

Now that the DPS contribution has been subtracted from the dihadron correlation,
we must include it separately to get an estimate of the $\Delta \varphi$-independent
background. The logarithmic divergence in the DPS limit is physically regulated
by confinement scale physics, and must be absorbed into a new nonperturbative
input describing the probe, namely the double parton distribution function
(DPDF).
For deuteron-gold collisions we divide the deuteron DPDF into two separate parts. 
The first one corresponds to taking two partons from the same nucleon in the 
deuteron,
described by a single-nucleon double parton distribution $D_{ij}(x_i,x_j,Q^2)$. 
We implement the kinematical constraint $x_i+x_j<1$ following
\re\cite{Strikman:2010bg} by taking
\begin{multline}
\label{eq:kc}
D^{(1)}_{ij}(x_i,x_j,Q^2) = \frac{1}{2} \bigg[f_i(x_i)f_{j}\left(\frac{x_j}{1-x_j}\right) 
\\
+ f_i\left( \frac{x_i}{1-x_j}\right) f_{j}(x_j) \bigg],
\end{multline}
 where $i$ and $j$ denote the parton species ($g,u,d$). 
For the deuteron one must sum over the 
proton and neutron:
 $D^d_{ij}(x_i,x_j,Q^2) = D^p_{ij}(x_i,x_j,Q^2) + D^n_{ij}(x_i,x_j,Q^2)$
The second contribution involves taking one parton from the neutron and 
the other one from the proton, which is not bound by the same kinematical
constraint. In this case one must sum; not symmetrize; over the different combinations 
of different partons from different nucleons:
\begin{equation}
\label{eq:nokc}
D^{(2)}_{ij}(x_i,x_j,Q^2) =  \left[f^p_i(x_i)f^n_{j}\left(x_j\right) 
+ f^p_i\left(x_i\right) f^n_{j}(x_j) \right].
\end{equation}
The double parton scattering contribution is the 
sum of these two
\begin{multline}
 \label{eq:totDPS}
\frac{\ud N^{h_a h_b \to ijX}_{\textrm{DPS}}}{\ud y_q \ud^2 \kt \ud y_g \ud^2 \qt }
= 
\frac{1}{(2\pi)^4}
\bigg(D^{(1)}_{ij}(x_i,x_j,Q^2) 
\\
+ D^{(2)}_{ij}(x_i,x_j,Q^2)\bigg) 
S_i(\kti)S_j(\ktj),
\end{multline}
where the two point functions $S_i(\kti),S_j(\ktj)$ are taken in the fundamental 
or adjoint representation according to the parton species $i,j$.

The total quark-gluon production cross section is now
the sum of~\nr{eq:totDPS} and~\nr{eq:dAuqg-xs}:
\begin{multline}
 \label{eq:totqglevel}
\ud N^{h_a h_b \to ijX}
=
\ud N^{h_a h_b \to ijX}_{\textrm{DPS}}
+ 
\ud N^{h_a h_b \to ijX}_{\textrm{sub}}.
\end{multline}

After convolution with fragmentation functions the cross section for
double inclusive production of hadrons $1$ and $2$ becomes
\begin{multline}
 \label{eq:tot}
\frac{\ud N^{h_a h_b \to h_1 h_2 X}}{\ud y_1 \ud^2 \pti \ud y_2 \ud^2 \ptii }
= 
\int \frac{\ud z_1}{z_1^2} \frac{\ud z_2}{z_2^2} 
\\
\frac{\ud N^{h_a h_b \to ijX}}{\ud y_1 \ud^2 (\pti  / z_1) \ud y_2 \ud^2 (\ptii/z_2) }
D_{h_1,i}(z_1)D_{h_2,j}(z_2) 
\end{multline}
where $i$ and $j$  are summed over gluons and light quark flavors; for the 
combination $i,j=q,g$ and $i,j=g,q$ the parton level cross section includes
the correlated cross section $\ud N^{h_a h_b \to ijX}_{\textrm{sub}}$ and
for the others only the DPS contribution.
This is our final expression that will 
be compared to the experimental yield. 
We shall now turn to 
the calculation of the different Wilson line correlators appearing in 
$\ud N^{h_a h_b \to ijX}_{\textrm{sub}}$.

\section{Multi-point correlators of Wilson lines} 
\label{sec:multipt}

As pointed out in \cite{Dumitru:2010mv,Dumitru:2010ak,Dumitru:2011vk},
one expects the true JIMWLK result for higher-point 
correlators of Wilson lines to significantly deviate from the
 simple product of dipole correlators (called ``naive large
$\nc$ limit'' in~\cite{Dumitru:2011vk}) used in the phenomenological 
applications so far~\cite{Tuchin:2009nf,Albacete:2010pg}.
It was conjectured in \re\cite{Dumitru:2011vk}, 
based on numerical tests for particular coordinate 
configurations, that JIMWLK evolution for
the higher-point correlators of Wilson lines can
be approximated by a ``Gaussian approximation''. 
Here one constructs, as in the MV model, the Wilson lines
from color charge densities whose probability distribution 
is assumed to be local in rapidity and Gaussian. This enables one 
to express all higher-point functions in terms of the two point function.
The equations relating different Wilson line correlators are
the same in the Gaussian approximation of JIMWLK and the MV model, the
difference between the two being the different functional form of the two 
point function (dipole). A theoretical explanation for the 
success of the Gaussian approximation is given in 
\re\cite{Iancu:2011ns,Iancu:2011nj}.

The cross section formula \nr{eq:dihadron-xs} 
involves correlators of up to six Wilson lines. 
The general Gaussian approximation for the six point function
$S^{(4)}$ is not known at finite~$\nc$,
 so we will here use the large~$\nc$ limit. We shall also compare
the large~$\nc$ limit to the approach used in  
\re\cite{Dumitru:2011vk} for the six point function, which we call here
the ``factorized Gaussian'' approximation.
This consists of expressing all the higher-point functions 
as products of \emph{fundamental} representation traces, and then assuming 
that the expectation value of a product of traces factorizes
into a product of expectation values. While this factorization 
would follow from the large $\nc$-limit, it is a consistent 
approximation scheme in itself also at finite $\nc$.  
The ``factorized Gaussian'' has the advantage of preserving
all the ``coincidence limits'', i.e. it correctly incorporates
the constraints coming from the SU($\nc$) group definition
$V(\xt)V^\dag(\yt)\to 1,$ when $\xt\to\yt$. 
At a minimum,
comparing this approximation to the large-$\nc$ result should
give us an indication of the size of finite $\nc$ effects. 
Evidence from a numerical calculation~\cite{Kovchegov:2008mk}
of two traces appearing on the r.h.s. of the BK equation
shows that, at least in this particular case, the ``factorized
Gaussian'' approximation is much more accurate than the 10\%~level
suggested by simple $\nc$ counting.
While for the four-point function 
$S^{(3)}$ it would be possible to use the full Gaussian approximation,
we will, for consistency, use the same approach also for $S^{(3)}$.
Thus the ``factorized Gaussian'' formulae for the Wilson line correlators used in 
\eq\nr{eq:dihadron-xs} are
\begin{widetext}
\begin{align}
\label{eq:gaussfin}
S^{(4)}(\bt,\btp,\xt,\xtp) &\approx
\frac{\nc^2}{\nc^2-1} 
\left[ S(\xt,\xtp) Q(\bt,\btp,\xt,\xtp) 
-\frac{1}{\nc^2}
S(\bt,\btp)
 \right]
\\
S^{(4)}_\textrm{DPS}(\bt,\btp,\xt,\xtp) &\approx
\frac{\nc^2}{\nc^2-1}S(\bt,\btp) 
\left[S(\xt,\xtp)^2 -\frac{1}{\nc^2}\right]
\\
S^{(3)}(\bt,\xt,\ztp) &\approx 
\frac{\nc^2}{\nc^2-1}
\left[S(\bt,\xt)S(\xt,\ztp)-\frac{1}{\nc^2} S(\bt,\ztp)\right]
\end{align}
\end{widetext}
where we use the  exact Gaussian quadrupole $Q(\bt,\btp,\xt,\xtp)$ calculated 
in \re\cite{Dominguez:2011wm}; since its expression is rather cumbersome 
we will not  repeat it here. 

In the large $\nc$ limit the Gaussian approximation reduces to 
\begin{widetext}
\begin{align}
\label{eq:infnc}
S^{(4)}(\bt,\btp,\xt,\xtp) & \underset{\nc\to\infty}{\approx}
S(\xt,\xtp) 
\Bigg[ S(\bt-\xt)S(\xtp-\btp)  \\
	&\quad - \frac{F(\bt,\xt,\xtp,\btp)}{F(\bt,\xtp,\xt,\btp)}
 \left( S(\bt-\xt)S(\xtp-\btp) - S(\bt-\btp)S(\xtp-\xt) \right) \Bigg],
\\
S^{(4)}_\textrm{DPS}(\bt,\btp,\xt,\xtp) &\underset{\nc\to\infty}{\approx}
S(\bt,\btp) S(\xt,\xtp)^2
\\
S^{(3)}(\bt,\xt,\ztp) &\underset{\nc\to\infty}{\approx} S(\bt,\xt)S(\xt,\ztp)
\end{align}
with the auxiliary function
\begin{equation}\label{eq:largencquad}
 \frac{F(\bt,\xt,\xtp,\btp)}{F(\bt,\xtp,\xt,\btp)} 
= \frac{ \ln S(\bt,\xtp) - \ln S(\bt,\btp) + \ln S(\xt,\btp) 
- \ln S(\xt,\xtp)}{\ln S(\bt,\xt) - \ln S(\bt,\btp) + \ln S(\xtp,\btp) 
- \ln S(\xtp,\xt) }.
\end{equation}
\end{widetext}
The first term in \eq\nr{eq:infnc} is the ``elastic'' term, which is the
only one kept in \re\cite{Marquet:2007vb}. The second term is sometimes
referred to as the ``inelastic'' one.
Note that one can explicitly verify that the limiting 
behavior $S^{(4)} \to  S^{(4)}_\textrm{DPS}$ holds both for the factorized Gaussian
approximation \nr{eq:gaussfin} and its large $\nc$ limit \nr{eq:infnc}.
The nonzero contribution in the DPS limit comes only from the ``inelastic'' part
of $S^{(4)}$ which was neglected in \re\cite{Marquet:2007vb}; thus the 
logarithmic divergence in this limit did not appear in that calculation.

\begin{figure}[tbp]
\begin{center}
\includegraphics[width=0.49\textwidth]{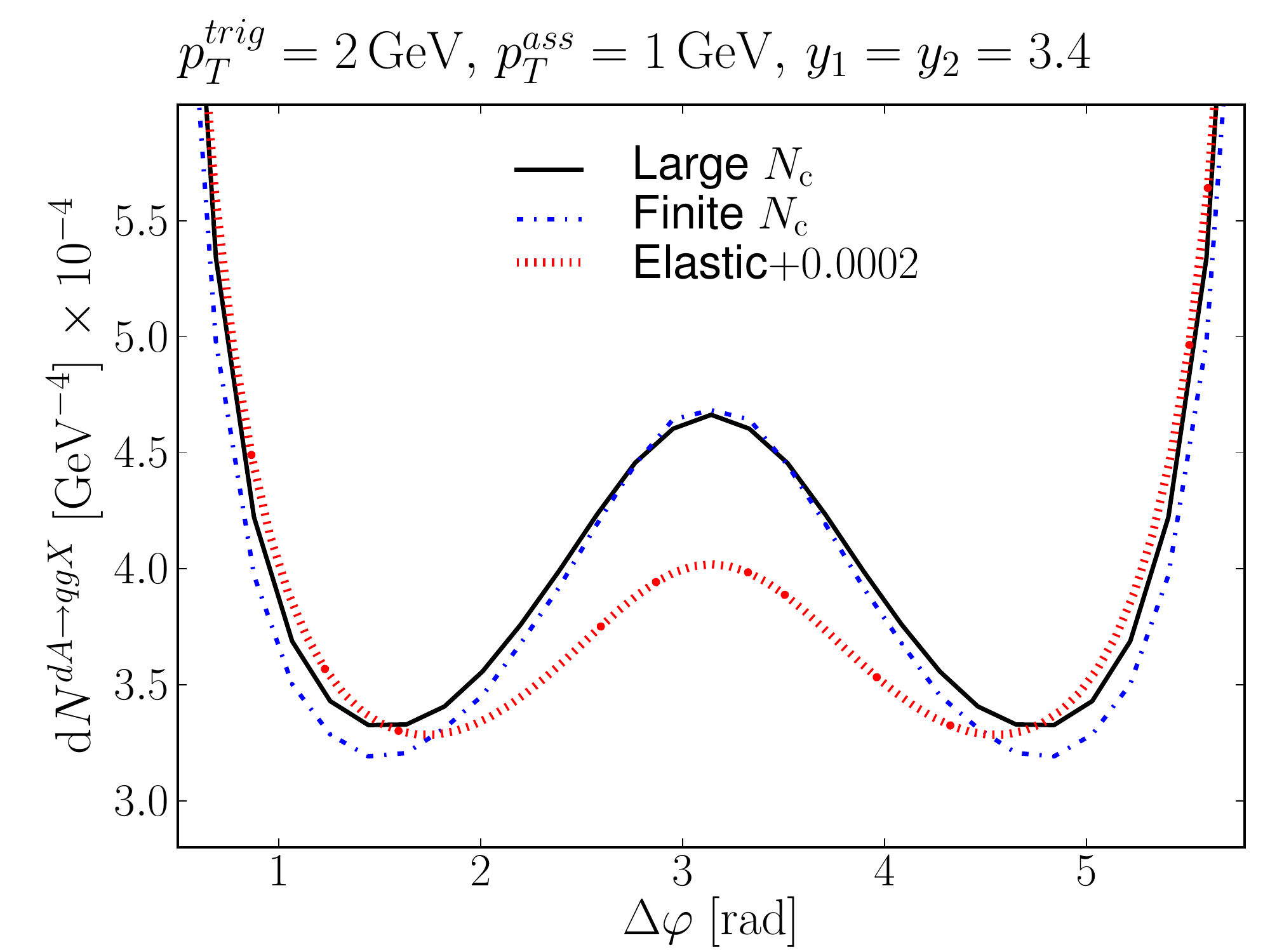}
\end{center}
\caption{
The quark-gluon  parton level azimuthal correlation in dAu collisions,
\eq \eqref{eq:dAuqg-xs},
near forward RHIC kinematics,
at transverse momenta $\ptrig=2\gev$, 
 $\pass=1\gev$ and $y=3.4$. Shown are the
``naive large $\nc$'' approximation with only the ``elastic'' contribution
used in \re\cite{Albacete:2010pg} and our Gaussian approximation 
of \eq\nr{eq:gaussfin} and its large $\nc$ limit \eq\nr{eq:infnc}. Note 
that a fixed $\Delta \varphi$-independent pedestal of $0.0002 \gev^{-4}$ has been added to
the ``elastic'' approximation for purposes of visualization. 
The DPS contribution is not included here.
} \label{fig:partonlev}
\end{figure}

\begin{figure}[tbp]
\begin{center}
\centerline{\includegraphics[width=0.49\textwidth]{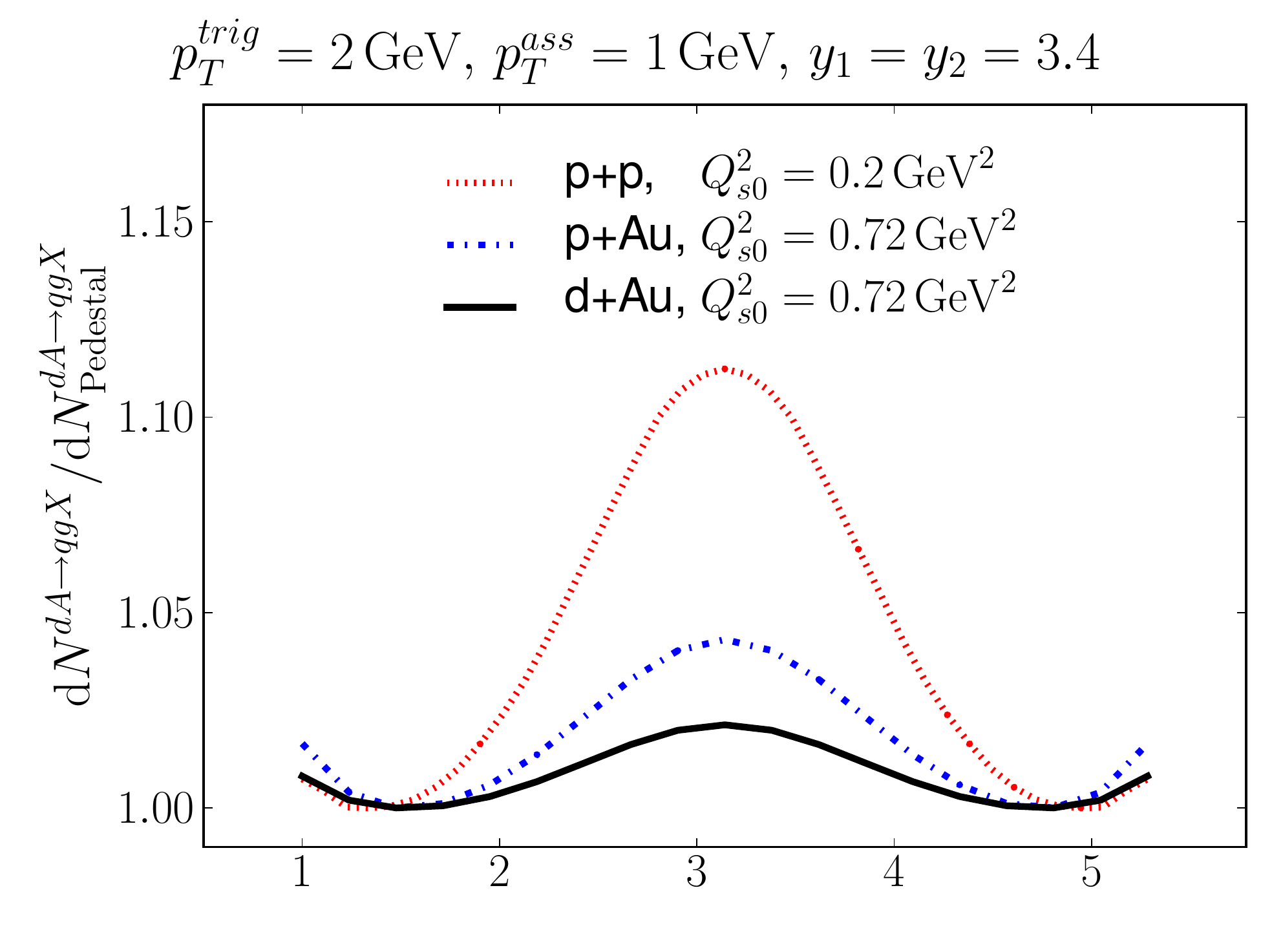}}
\centerline{\includegraphics[width=0.49\textwidth]{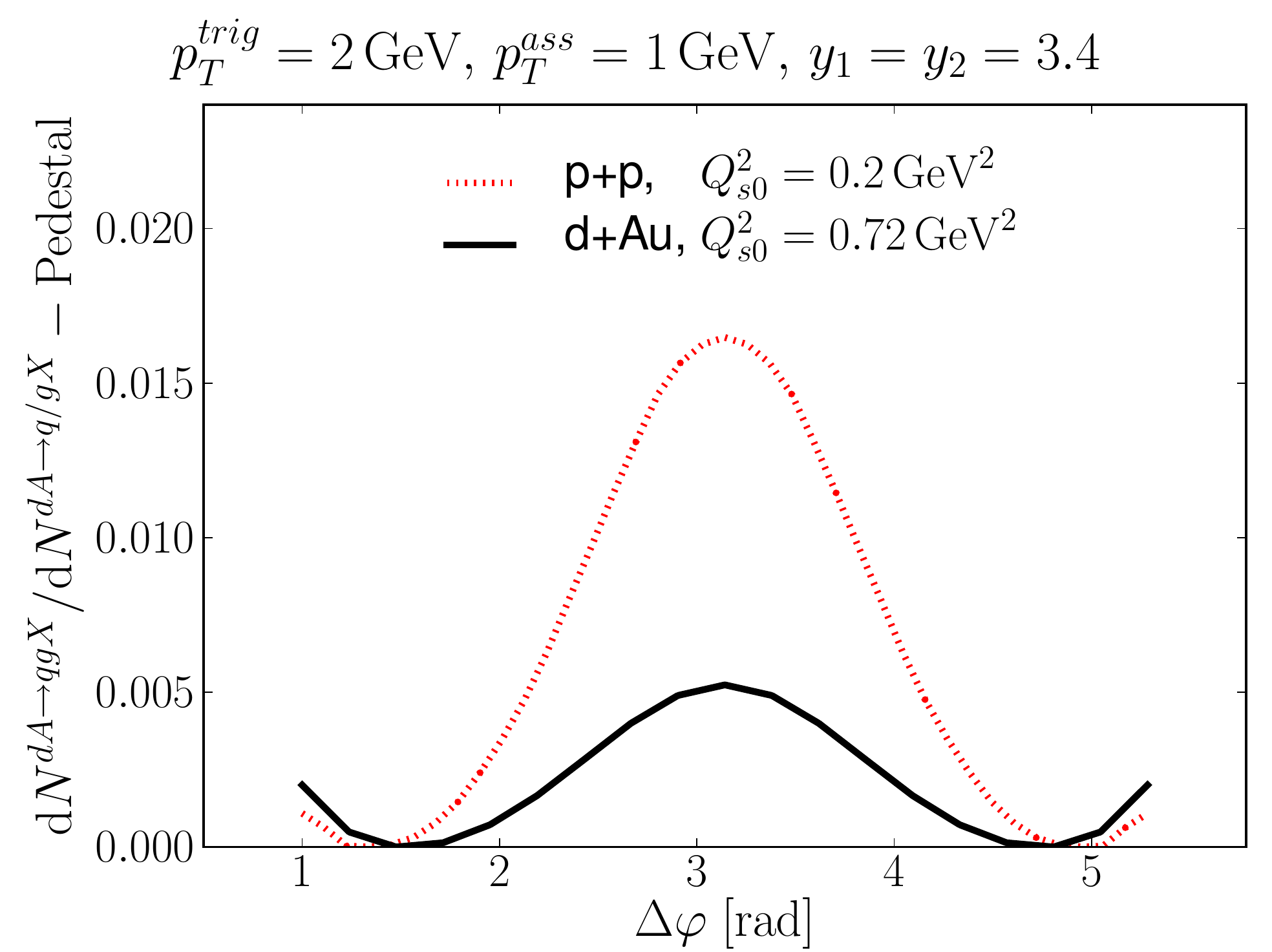}}
\end{center}
\caption{
The quark-gluon  parton level azimuthal correlation at forward RHIC kinematics
in proton-proton, proton-nucleus and deuteron-nucleus
collisions. 
Top:  normalized by the pedestal contribution,
bottom: pedestal contribution subtracted (in which case the pAu and dAu
results are identical by construction).
Shown is only the large $\nc$ result.
} \label{fig:partonlevvaryqs}
\end{figure}

\begin{figure}[tbp]
\begin{center}
\includegraphics[width=0.49\textwidth]{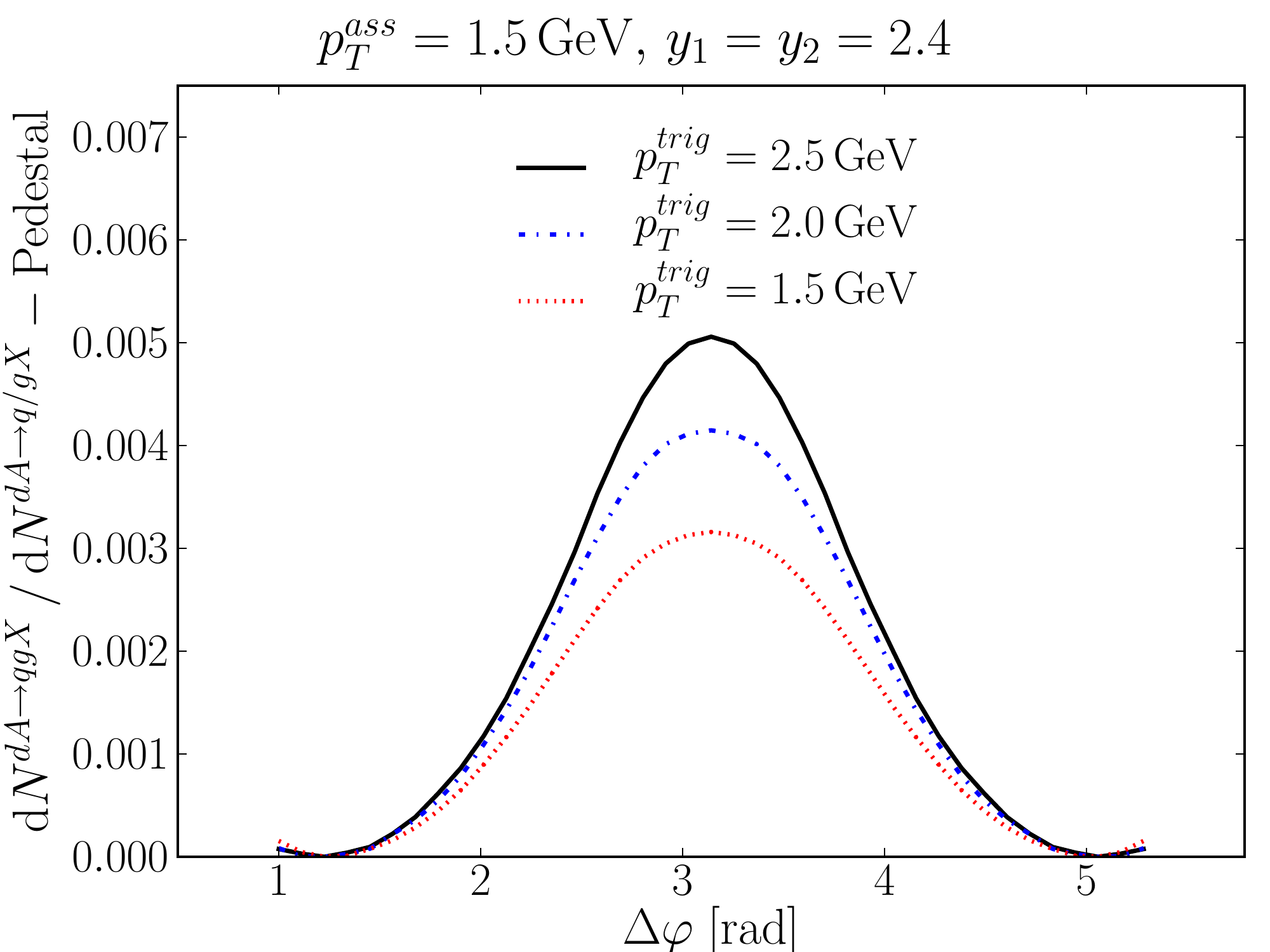}
\end{center}
\caption{
The pedestal-subtracted quark-gluon  parton level azimuthal correlation at 
forward RHIC kinematics,
at different values of the trigger transverse momenta $\ptrig=1.5,2,2.5\gev$
and $y=2.4$.
Shown is only the large $\nc$ result. 
} \label{fig:partonlevvarytrig}
\end{figure}

\begin{figure}[tbp]
\begin{center}
\includegraphics[width=0.49\textwidth]{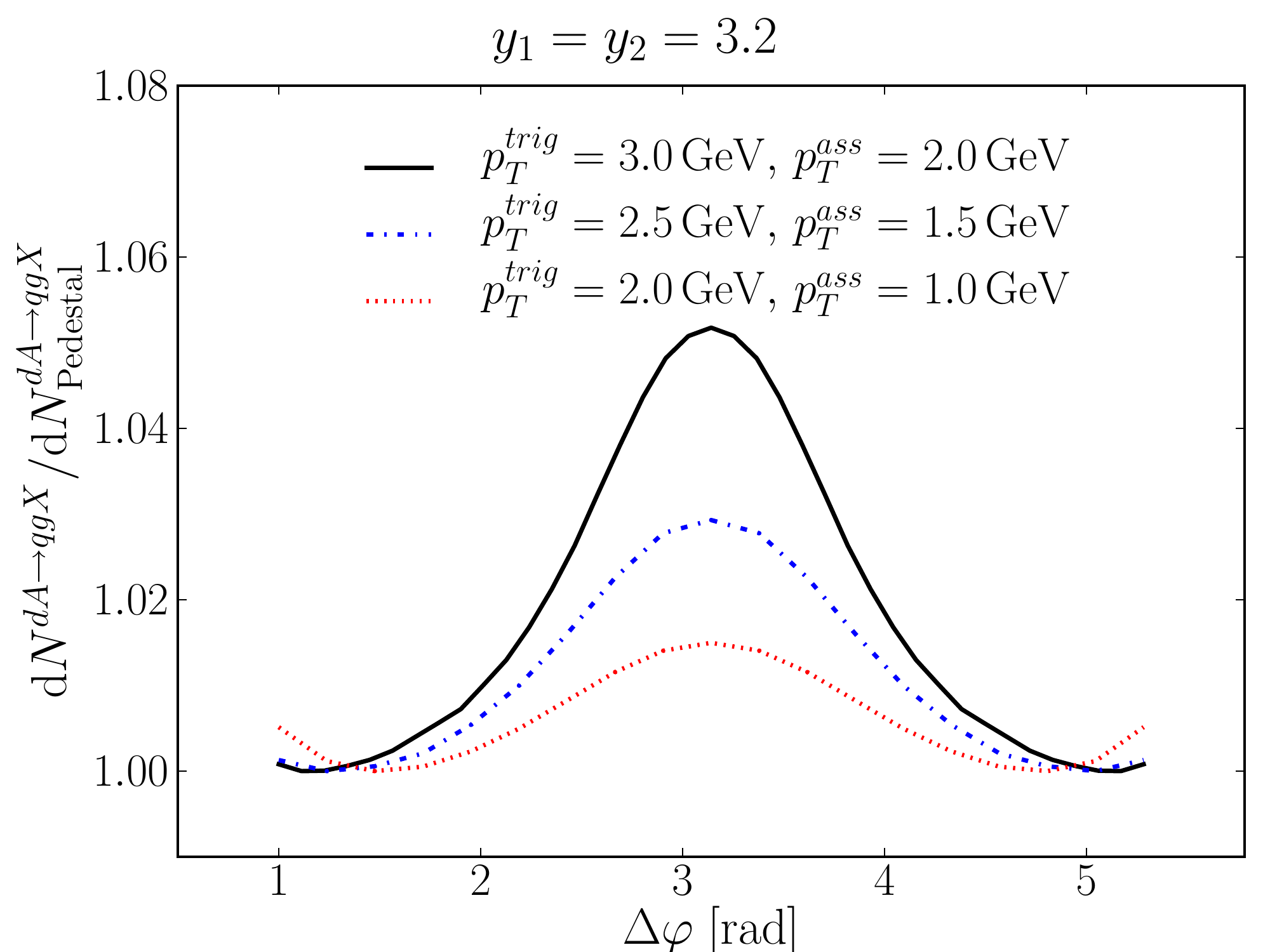}
\end{center}
\caption{
The total quark-gluon parton level dihadron production yield at forward RHIC kinematics
at different values of the trigger and associated particle transverse momenta
at $y=3.2$.
The DPS contribution is included and the result is normalized by the pedestal yield.
Shown is only the large $\nc$ result. 
} \label{fig:partonlevvarytrigass}
\end{figure}

\begin{figure}[tbp]
\begin{center}
\includegraphics[width=0.49\textwidth]{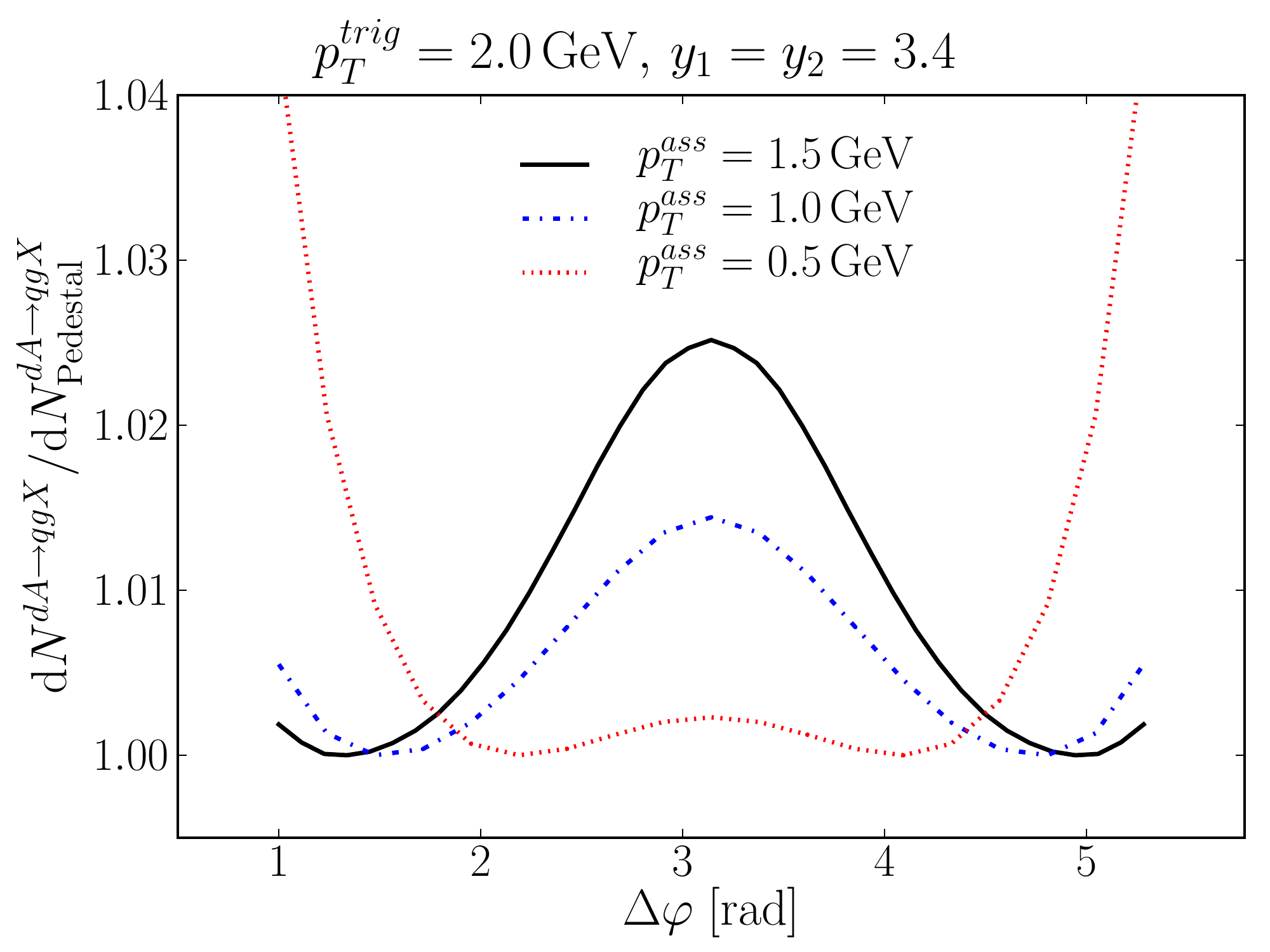}
\end{center}
\caption{
The total quark-gluon parton level dihadron production yield at forward RHIC kinematics,
for trigger transverse momentum $\ptrig=2\gev$, 
and different values of the associate
transverse momentum $\pass=0.5, 1,1.5\gev$ and $y=3.4$. 
Shown is only the large $\nc$ result.
The DPS contribution is included and the result is normalized by the pedestal yield.
} \label{fig:partonlevvaryass}
\end{figure}

\section{Results}
\label{sec:res}

\begin{figure}[tb]
\begin{center}
\includegraphics[width=0.49\textwidth]{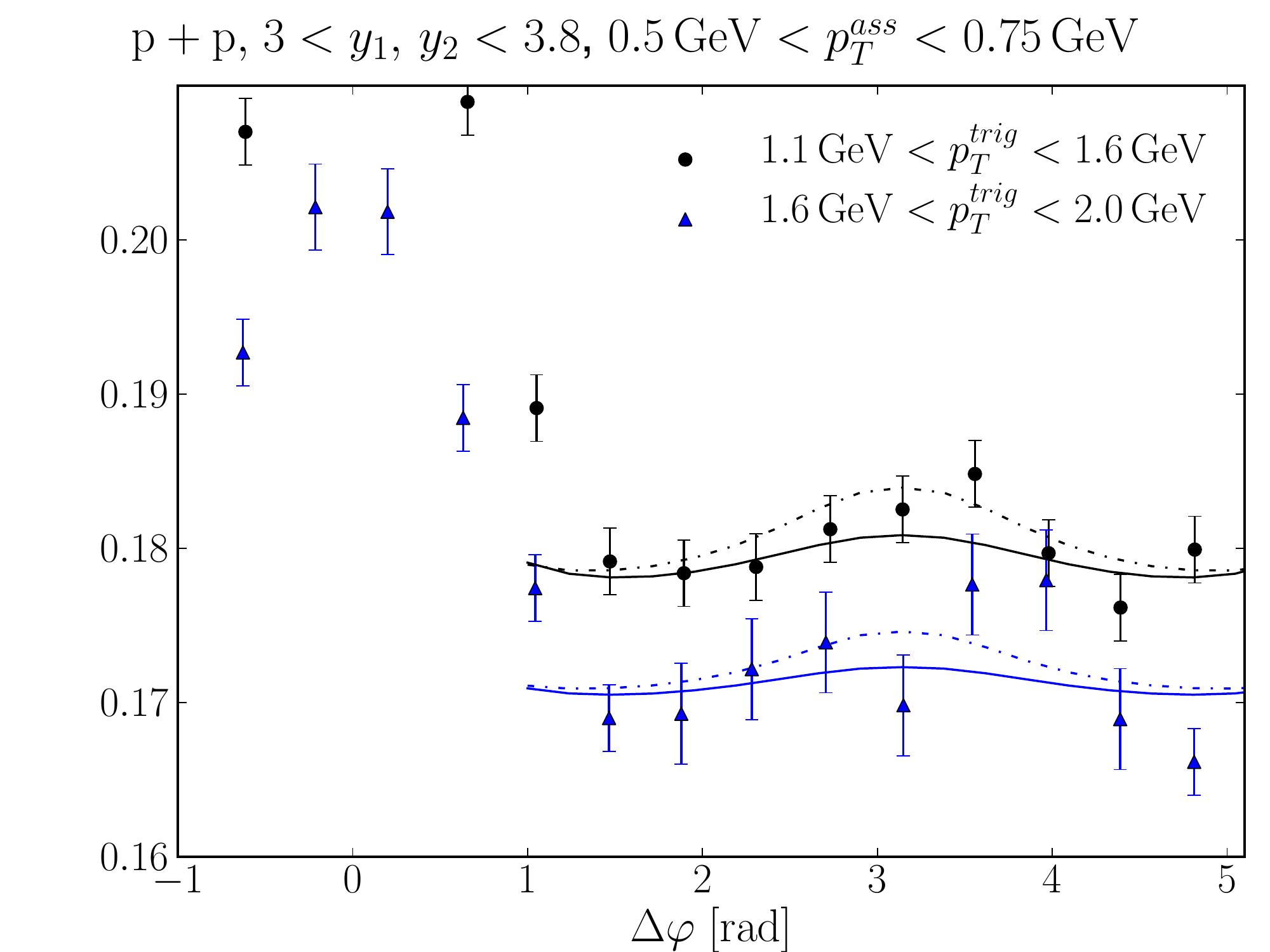}
\end{center}
\caption{
The $\pi^0$ azimuthal correlation compared to the 
PHENIX~\cite{Adare:2011sc} dAu result for two different $\ptrig$ bins. 
The $\Delta \varphi$-independent pedestal in the plot is adjusted to fit the
experimental data, see  Table~\ref{tab:ped} for the calculated estimates. 
The initial saturation scales are $\qso^2=1.51 \gev^2$ (solid line) and
$\qso^2=0.72 \gev^2$ (dashed line).
}\label{fig:compphenix}
\end{figure}

\begin{figure}[tb]
\begin{center}
\includegraphics[width=0.49\textwidth]{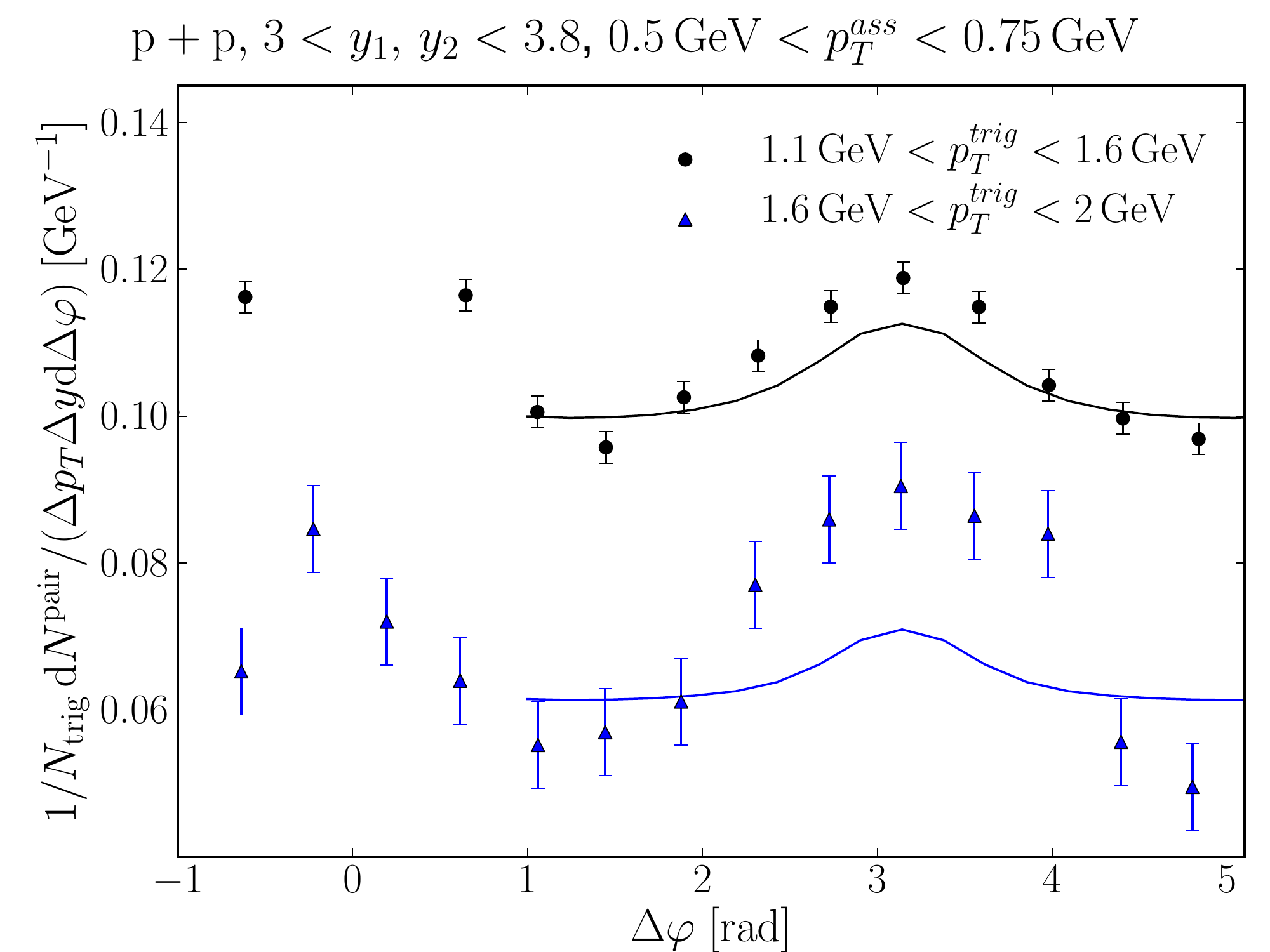}
\end{center}
\caption{
The $\pi^0$ azimuthal correlation compared to the 
PHENIX~\cite{Adare:2011sc} pp result, for the same 
transverse momenta as in \fig\ref{fig:compphenix}.
The $\Delta \varphi$-independent pedestal in the plot is adjusted to fit the
experimental data, see  Table~\ref{tab:ped} for the calculated estimates. 
The initial condition is the MV model with initial saturation 
scale $\qso^2=0.2 \gev^2$. Data points at larger $\ptrig$ are sifted
by $-0.03\gev^{-1}$ to avoid overlap.
}
\label{fig:compphenixpp}
\end{figure}

\begin{figure}[tb]
\begin{center}
\includegraphics[width=0.49\textwidth]{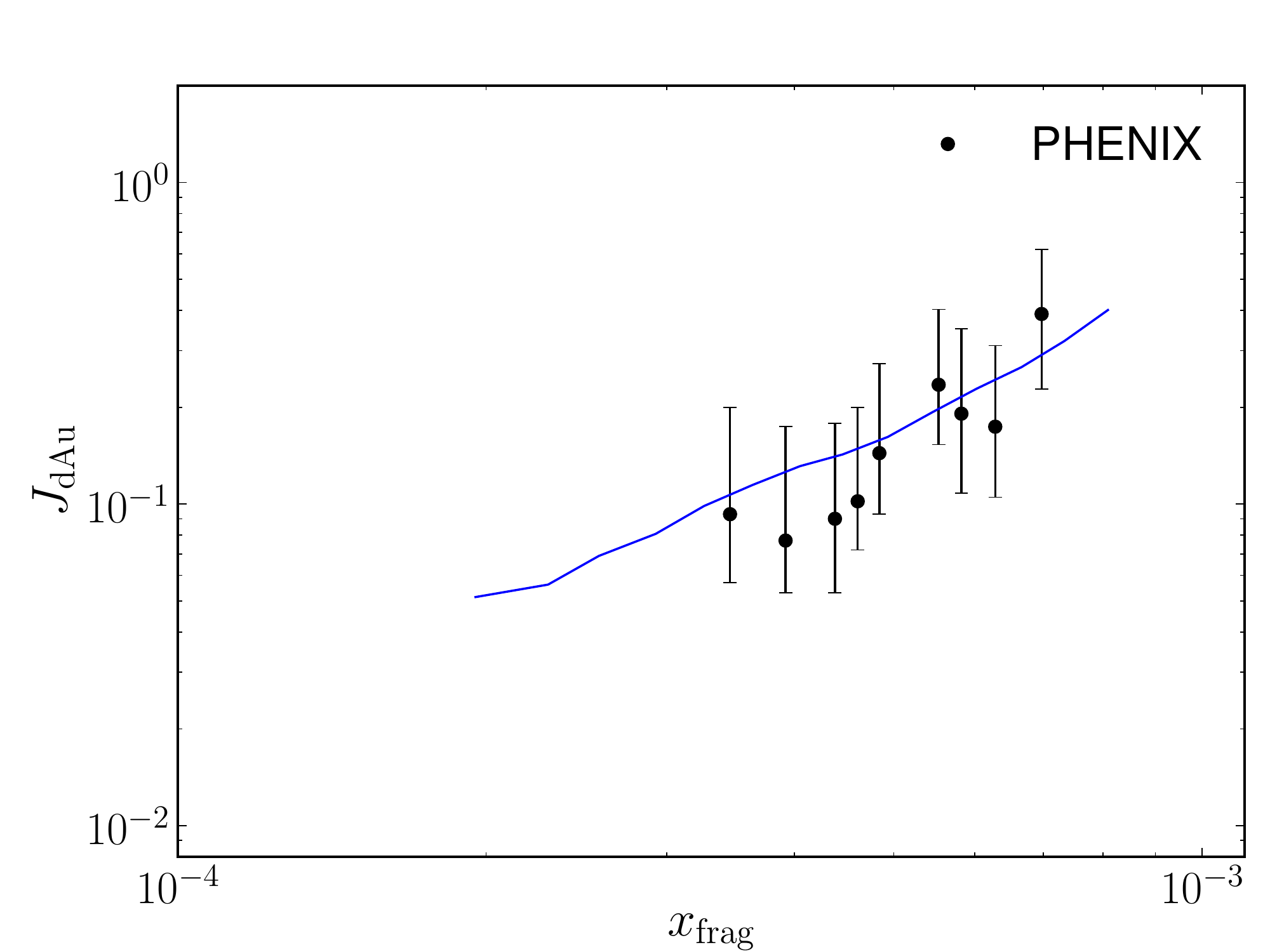}
\end{center}
\caption{
Integrated yield under the away side peak in central dAu collision
divided by the corresponding yield in pp compared
to the forward rapidity part of the PHENIX data~\cite{Adare:2011sc}. 
}
\label{fig:phenixjdau}
\end{figure}

\begin{figure}[tb]
\begin{center}
\includegraphics[width=0.49\textwidth]{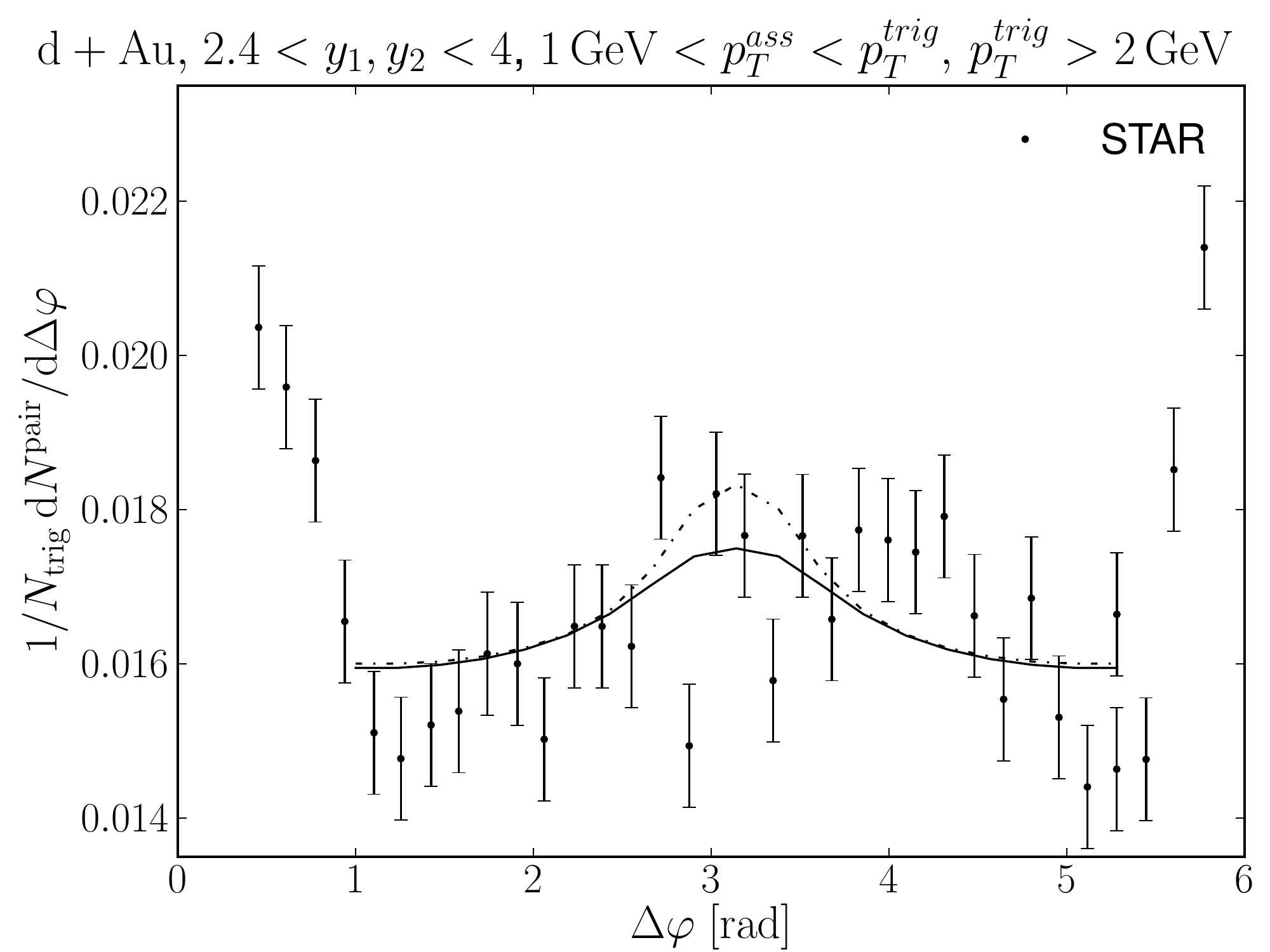}
\end{center}
\caption{
The $\pi^0$ azimuthal correlation compared to the 
preliminary STAR~\cite{Braidot:2011zj} result. 
The initial saturation scales are $\qso^2=1.51 \gev^2$ (solid line) and
$\qso^2=0.72 \gev^2$ (dashed line). 
}\label{fig:compstar}
\end{figure}

We begin this section by showing results at the parton level,
demonstrating the systematics as a function
of different trigger and associate momenta, different probe and target 
species. We then include the fragmentation functions for a 
more realistic comparison with both 
STAR and PHENIX dipion correlations.

Figure~\ref{fig:partonlev} shows the effect of the ``inelastic'' contribution 
neglected in \re\cite{Albacete:2010pg} to the dihadron correlation.
Keeping only the ``elastic'' term, as done in \re\cite{Albacete:2010pg},
results in an away side peak that is smaller by a factor of $\sim$2 than 
the factorized Gaussian approximation, while the effect on the width of the peak is 
much smaller. Our conclusion from this plot is that
including both contributions is essential for a meaningful 
comparison with experimental data.
We also notice that the finite~$\nc$
corrections to the Gaussian approximation can be neglected, and from now on
we only use the large~$\nc$ version of the Gaussian approximation.
The initial 
saturation scale for the nucleus is taken as $\qso^2=0.72\gev^2$.

Figure~\ref{fig:partonlevvaryqs} illustrates the difference 
between deuteron-nucleus, proton-nucleus and proton-proton collisions
in the ratio of the peak to the $\Delta \varphi$-independent background.
Plotted on top is the total dihadron cross section, including the DPS
contribution as discussed in \se\ref{sec:dps}, \emph{divided} by the 
minimum of the $\Delta \varphi$-distribution. On the bottom this
pedestal is \emph{subtracted}.
 The initial saturation scales are taken as 
 $\qso=0.2\gev^2$ for the proton and $\qso^2=0.72 \gev^2$ for a nucleus.
 Comparing the pp and pA/dA results 
one sees clearly the depletion of the peak relative to pp collisions
when the target is a large nucleus. This is the crucial generic feature
whose observation in the experimental data supports the concept 
of a depletion caused by intrinsic gluon transverse momenta in the
target given by a saturation scale.
 Since we are neglecting any differences
in the large~$x$ quark distribution between  a deuteron  and a proton,
there is no difference in the correlated peak between pAu and dAu
collisions. Due to the enhanced DPS contribution in deuteron collisions
(where the possibility of taking one large~$x$ quark 
from the proton and the other from the neutron
makes it much easier to observe a double scattering event in very
forward kinematics), the DPS contribution is much larger in
dAu collisions. This leads to a smaller ratio of the correlated
peak to the pedestal, purely because of the increase in the 
denominator. The effect is similar to what is seen in the STAR forward
neutron tagged (effectively pAu) data~\cite{Perkins:2011uc} and 
discussed in \re\cite{Strikman:2010bg}.

 Figure~\ref{fig:partonlevvarytrig} shows the dependence on
the parton level dAu result on the trigger transverse momentum.
Shown is the pedestal-subtracted yield per trigger for associate
parton transverse momentum $\ptt= 1.5\gev$ and trigger transverse
momenta $\ptt=1.5,\ 2.0, \  2.5\gev$. These should be compared
to the typical nuclear saturation scale $\qs\approx 1.5\gev$ at these 
rapidities (defining $\qs$ via the saturation criterion $S(r=1/\qs)=1-e^{-1/4}$).
One again sees clearly the expected effect of the nuclear saturation scale:
the peak is small for small trigger 
transverse momentum, but starts to reappear when the trigger momentum is
increased. Note that the rapidity in this plot is slightly less 
forward than elsewhere. This is to avoid hitting the kinematical limit 
$x_g+x_q<1$ at large $x$  in the deuteron. At very forward rapidites
and high enough transverse momenta the kinematical limit causes the
height of the peak in the yield per
trigger to decrease when the trigger $\ptt$ is increased. This 
is caused by the fact that the correlated quark-gluon pair in the numerator
of the ratio requires a larger total longitudinal momentum (and is thus 
suppressed at $x\to1$) than the single quark (at the same $y$ and $\ptt$ as
the trigger) in the single inclusive cross section in the denominator.
Because of this effect  the dependence of the
peak height on the trigger $\ptt$ at forward RHIC kinematics is actually 
smaller than the ratio of the momentum to the saturation scale would
suggest.

Figures~\ref{fig:partonlevvarytrigass} and~\ref{fig:partonlevvaryass}
show the effect on the yield per trigger of varying also the associate 
$\ptt$. Since the two particle yield (just like the single inclusive associate
particle yield) is a steeply falling function of the associate 
$\ptt$, the yield per trigger itself falls as a function of $\pass$.
 To show what
is actually happening to the peak with respect to the background
we again include the DPS contribution to the pedestal 
and divide the correlation by the $\Delta \varphi$-independent part.
 As expected, the away-side peak is strongly suppressed for 
smaller trigger $\ptt$. 

\begin{table*}
\begin{tabular}{|l|l||r|r|}
\hline
Data & $\ptt$ range & pedestal & exp.  \\
\hline\hline
PHENIX pp & $1.1\gev  < \ptrig < 1.6\gev$ &  0.04 & 0.095\\ 
\hline
PHENIX pp & $1.6\gev  < \ptrig < 2.0\gev$ &  0.02 & 0.079\\ 
\hline
PHENIX dAu & $1.1\gev  < \ptrig < 1.6\gev$ &  0.10 & 0.176\\ 
\hline
PHENIX dAu & $1.6\gev  < \ptrig < 2.0\gev$ &  0.08 & 0.163\\ 
\hline
STAR dAu & $2\gev  < \ptrig,\ 1\gev < \pass < \ptrig$ &  0.02 & 0.0145\\ 
\hline
\end{tabular}
\caption{Calculated estimates for the pedestal height compared to
the experimental values. In PHENIX results the units are $\gev^{-1}$.
The dAu values are for central
collisions, and $\qso^2$ is taken as $\qso^2=1.51\gev^2$.
}
\label{tab:ped}
\end{table*}

Figures~\ref{fig:compphenix} and \ref{fig:compphenixpp}
compare our calculation of the away side correlation to the 
PHENIX~\cite{Adare:2011sc}  data for different momentum bins. 
Figure~\ref{fig:compphenixpp} shows the proton-proton  and 
\ref{fig:compphenix} the central deuteron-gold result. For the latter we
show calculations using two initial values for the saturation scale:
$\qso^2=1.51\gev^2$, which would be the natural estimate for these
central collisions, and a smaller value $\qs^2= 0.72\gev^2$, 
which would be preferred by some of the single inclusive data.

Due to the significant uncertainty in the single inclusive baseline spectrum,
the $\Delta \varphi$-independent pedestal values are rough 
estimates that cannot be directly compared to the data at this point. 
For plotting purposes the pedestal values have been adjusted to the data
in the plot;
the results of the calculation compared to the experimental pedestal 
are given in Table.\ref{tab:ped}.
The normalization uncertainty also affects the height of the away side peak,
even though we would expect that at least
part of this theoretical uncertainty cancels in the yield-per-trigger ratio.
Taking into account these uncertainties, we obtain relatively good description
of the back-to-back peak, underestimating the peak height especially 
with $\qso^2=1.51\gev^2$ for the nucleus and in the case of the proton-proton
scattering.

PHENIX has also published data for larger trigger particle momentum $2 < \ptrig < 5 \gev$. 
We can not describe the observed relative increase of the peak height compared
to lower values of the trigger momenta in this kinematics. This phenomenon can 
be seen in parton level results (see \fig \ref{fig:partonlevvarytrig}), but
in PHENIX kinematics we are so close to the kinematical boundary that
in our calculation the 
deuteron parton distribution function suppresses the peak.

Figure~\ref{fig:phenixjdau} compares our result for the nuclear modification
factor $J_{\textrm{dAu}}$
of the area under the peak to the PHENIX experimental result~\cite{Adare:2011sc}.
The experimental data are given as a function of 
$x_\textrm{frag} \equiv 
\left( \langle \ptrig \rangle  
e^{- \langle \eta_\textrm{trig}\rangle }
+
\langle \pass \rangle 
e^{- \langle \eta_\textrm{ass}\rangle }\right)/\sqrt{s_{NN}}.
$
We compute this quantity by calculating the ratios of the areas
under the pp and dAu away side peaks at various kinematical configurations
at rapidities $3<y_1,y_2<3.8$ and at various trigger ($1<\ptrig<2\gev$) and
associate particle ($0.5<\pass<1.5\gev$) transverse momenta. We then obtain
$J_{\textrm{dAu}}$ by averaging the results in every $x_\textrm{frag}$ bin.
This is not exactly the same method that is used in experiments, but 
we have tested that the result does not depend much on the kinematical limits.
Our calculation agrees 
with the PHENIX result within errors, and we clearly see that
the suppression increases when $x_{\mathrm{frag}}$ decreases, which is
expected as the saturation effects should become more visible at smaller
$x_{\mathrm{frag}}$. 
The peak height is underestimated
in both pp and dAu collisions probably due to the uncertainties in the
single inclusive baseline normalization, and we expect that 
this error partly cancels in $J_{\textrm{dAu}}$. 

Finally \fig\ref{fig:compstar} compares our calculation 
to the STAR~\cite{Braidot:2011zj} result. While the errors in the data
are rather large, let us point out a few things. Our calculation
seems to agree reasonably well with the data, although perhaps
underestimating the hight of the peak. It is clear from
\fig\ref{fig:partonlev} that neglecting the ``inelastic'' part of the 
cross section (as done by Albacete and Marquet~\cite{Albacete:2010pg})
would make the agreement worse.\footnote{
When comparing to the result in \re\cite{Albacete:2010pg} note
 also that there is a significant numerical error in the calculation
of \re\cite{Albacete:2010pg},  which we thank C. Marquet for 
indicating to us.} Note also that we are assuming collinear factorization
of partons into hadrons, which is very crude at these small transverse
momenta. Any $\ktt$-smearing from fragmentation would broaden the 
away-side peak.

\section{Conclusions}

We have in this paper performed a calculation of nuclear modifications
of forward dihadron correlations in the CGC framework. 
We use a  running coupling BK evolution for the dipole and a factorized
Gaussian approximation for the higher-point functions
of Wilson lines to describe the target.
Our calculation is
the first one in the literature to evaluate the
the full expression of the dihadron cross
section, keeping both the ``elastic'' and ``inelastic'' terms
(unlike  \re\cite{Albacete:2010pg}) and
not restricted to the high-$\ptt$ ``correlation limit''
(unlike  \re\cite{Stasto:2011ru}).
We find that including the ``inelastic'' term 
in  has a significant effect on the correlation,
enhancing the peak by a factor $\sim$2. We also show that in the 
appropriate kinematics the dihadron cross section reduces to  
$\Delta \varphi$-independent double parton scattering, 
which in the earlier literature has been considered
as a completely separate contribution.

We show explicitly how the large saturation scale in a nucleus
leads to a smoothing of the away-side peak in dihadron correlations
for transverse momenta of the order of the saturation scale.
We obtain a reasonable order-of-magnitude estimate also for the
$\Delta \varphi$-independent pedestal part of the correlation. 
Taking into account the lack of a parametrization of
the dipole cross section that would simultaneously reproduce both
 mid- and forward rapidity 
single inclusive hadron spectra from all
the RHIC experiments (including the STAR and PHENIX forward 
$\pi^0$ data sets for which the correlation measurements are performed)
it is difficult to consistently perform a more accurate calculation
of the pedestal contribution at this point. We thus leave
a more thorough exploration of the single inclusive spectra for 
future work. Results from the future LHC proton-nucleus run,
with the significantly larger kinematical coverage available,
should significantly clarify these uncertainties. A realistic 
treatment of the LHC kinematics will require taking into account
also the gluon-initiated channel, which has not been done yet in this
work.

\section*{Acknowledgements}
We thank K.~J.~Eskola, I.~Helenius, R.~Paatelainen,
B.~Schenke, M.~Strikman and R.~Venugopalan for discussions 
and J.~Albacete and C.~Marquet for helpful
comparisons with their results.
H.M. is supported by the Graduate School of Particle and Nuclear Physics.
This work has been supported by the Academy of Finland, projects 141555 
and 133005, and by computing resources from
CSC -- IT Center for Science in Espoo, Finland. 

\appendix

\section{$S^{(4)}$ and $S^{(2)}$ in the DPS limit}\label{app:dps}

\begin{figure}[tb]
\begin{center}
\includegraphics[width=0.49\textwidth]{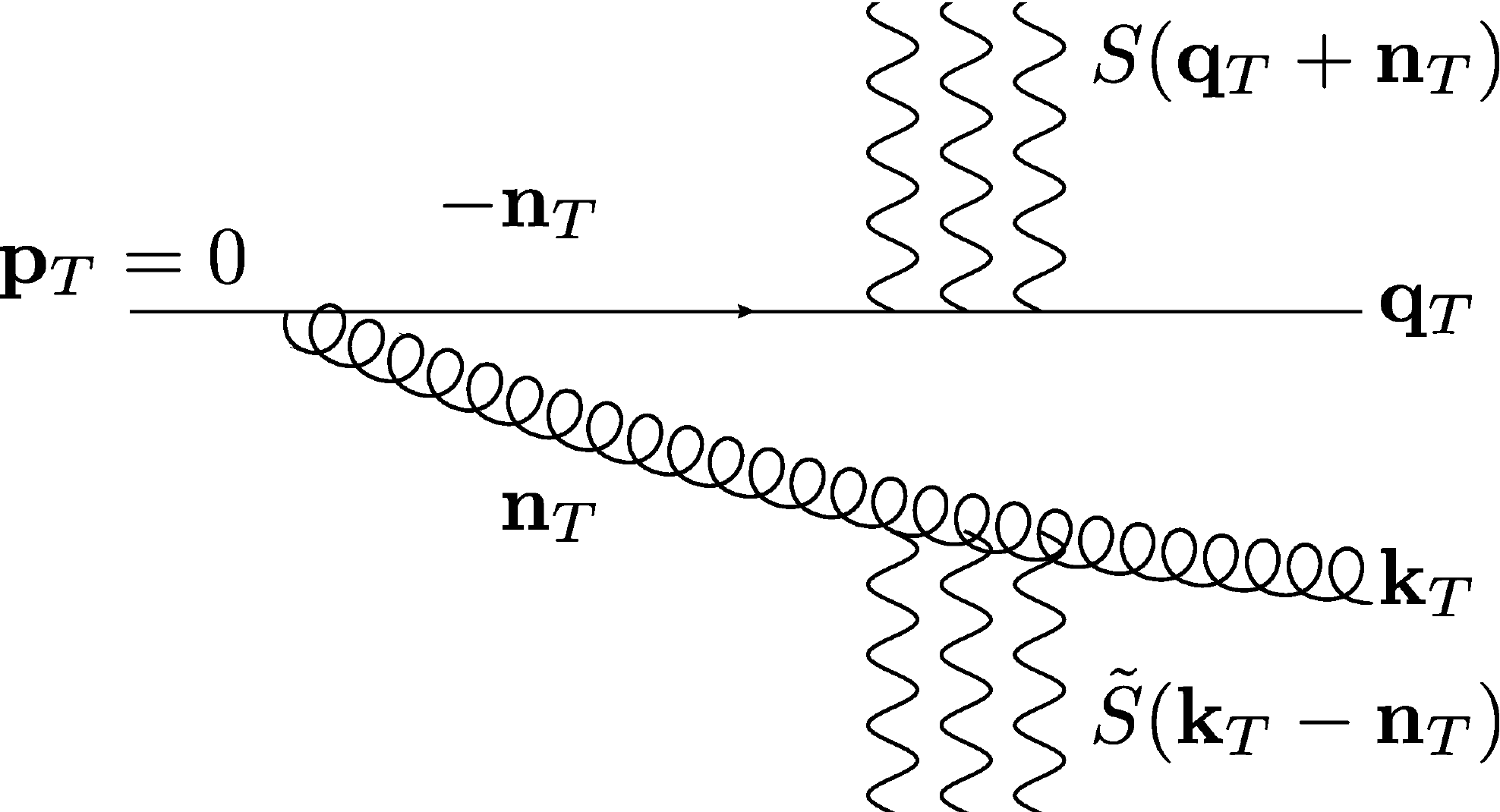}
\end{center}
\caption{
Illustration of the kinematics in \eq\nr{eq:dihadron-xs-dps3}.
}\label{fig:dpskin}
\end{figure}

Let us now show that only the four point function $S^{(4)}$ gives an infrared
divergent contribution to the the double inclusive cross section. 
We then show how this contribution can be identified as the double 
parton scattering contribution appearing in the literature, 
although calculated in the ``hybrid formalism'' and not collinearly factorized
perturbation theory as usual.

In the DPS limit  and for massless quarks the double inclusive cross section 
\eq\nr{eq:dihadron-xs} reduces to 
\begin{widetext}
\begin{multline}
\label{eq:dihadron-xs-dps}
\frac{\ud \sigma^{hA\to qgX}}{\ud y_q \ud y_g \ud^2 \qt  \ud^2 \kt  } 
\underset{DPS}{=} x q_h(x) z (1-z) \as 
\int \frac{\ud^2 \xt}{(2\pi)^2} \frac{\ud^2 \xtp}{(2\pi)^2} 
\frac{\ud^2 \bt}{(2\pi)^2} \frac{\ud^2 \btp}{(2\pi)^2} 
e^{i \kt \cdot(\xtp - \xt)} e^{i\qt\cdot(\btp-\bt)} 
8 \pi^2 \frac{\ut \cdot \utp}{|\ut|^2|\utp|^2}
\\
P_{q\to qg}(z)
\left[
S(\bt-\btp) \widetilde{S}(\xt-\xtp) 
+ S(z(\xt-\xtp) + (1-z)(\bt-\btp))
\right],
\end{multline}
\end{widetext}
where we recall that $ \widetilde{S}$ denotes the adjoint representation two point 
function and that $\ut= \xt-\bt; \  \utp= \xtp-\btp$. 
We have denoted the splitting
function for radiating a gluon from a quark by
$P_{q\to qg}(z) = \cf (1+(1-z)^2)/z$.
Using the momentum representation
\begin{equation}
2 \pi \frac{\ut}{\ut^2} = -i \int \ud^2 \nt e^{i \nt \cdot \ut}\frac{\nt}{\nt^2},
\end{equation}
shifting the integration variables $\xt$ and $\bt$ by $\xtp$ and $\btp$ 
respectively,
then integrating over $\xtp$ and $\btp$
we get
\begin{widetext}
\begin{multline}
\label{eq:dihadron-xs-dps2}
\frac{\ud \sigma^{hA\to qgX}}{\ud y_q \ud y_g \ud^2 \qt  \ud^2 \kt  } 
\underset{DPS}{=} x q_h(x) z (1-z) 
\as S_\perp
\frac{1}{(2\pi)^6}
\int \ud^2 \xt \ud^2 \bt
\frac{\ud^2 \nt }{\nt^2} 2
e^{- i (\kt-\nt) \cdot  \xt } e^{- i(\qt+\nt)\cdot\bt} 
\\
P_{q\to qg}(z)
\left[
S(\bt) \widetilde{S}(\xt) 
+ S(z\xt + (1-z)\bt)
\right],
\end{multline}
\end{widetext}

Let us first look at the second term, which originates in $S^{(2)}$. 
Changing the integration variable from $\bt$ to $\zt = z\xt+(1-z)\bt$ the
$\zt$-integral gives a two-point function in momentum space and the $\xt$-integral a
$\delta$-function which can be used to integrate over $\nt$ to get 
\begin{equation}
 x q_h(x) z (1-z) \as S_\perp
\frac{1}{(2\pi)^4}
2 P_{q\to qg}(z) \frac{ S(\qt+\kt)  }{[ (1-z)\kt - z \qt]^2}.
\end{equation}
This is a finite result, which is included as such in \eq\nr{eq:dAuqg-xs}.

The first term of \eq\nr{eq:dihadron-xs-dps2}, on the other hand, becomes
\begin{multline}
\label{eq:dihadron-xs-dps3}
\frac{\ud \sigma^{hA\to qgX}}{\ud y_q \ud y_g \ud^2 \qt  \ud^2 \kt  } 
\underset{DPS}{=} x q_h(x) z (1-z) 
P_{q\to qg}(z)
\\
\as S_\perp
\frac{1}{(2\pi)^6}
\int \frac{\ud^2 \nt }{\nt^2} 2
S(\qt+\nt) \widetilde{S}(\kt-\nt).
\end{multline}
The interpretation of this result is illustrated in \ref{fig:dpskin}: the 
emission a gluon with transverse momentum $\nt$ is followed by the subsequent 
independent scattering of the quark and the gluon off the target.
The integral over $\nt$ is logarithmically divergent in the infrared $\ntt \to 0$. 
Regulating it at a parametrically confinement scale momentum $\Lambda$ we can
neglect $\nt$ compared to $\kt,\qt$ in the two point functions and approximate
\begin{multline}
\label{eq:dihadron-xs-dps4}
\frac{\ud \sigma^{hA\to qgX}}{\ud y_q \ud y_g \ud^2 \qt  \ud^2 \kt  } 
\underset{DPS}{=} 
x q_h(x) z (1-z)  P_{q\to qg}(z) 
\\
\frac{\as}{2 \pi} \ln \Lambda^2
\frac{1}{(2\pi)^4}
S(\qt) \widetilde{S}(\kt).
\end{multline}
We find it instructive to compare this
 to the general form for the LO~DGLAP evolution 
equation for the quark-gluon double parton distribution
$D_{qg}(x_q,x_g,Q^2)$ (see e.g. \eq(2.1) of \re\cite{Gaunt:2009re}):
\begin{multline}
\frac{\ud D_{qg}(x_q,x_g,Q^2) }{\ud \ln Q^2}
= \frac{\as(Q^2)}{2 \pi} \bigg[\textnormal{evol.}
+ \\
q(x_g+x_g,Q^2)\frac{1}{x_q+x_g}P_{q\to qg}\left(\frac{x_g}{x_q+x_g}\right)
\bigg].
\end{multline}
Here ``evol.'' denotes DGLAP evolution terms corresponding to separate 
radiation from quarks and gluons, leading to a variation of
the double parton distribution with the scale $Q^2$. As we are 
not interested in very high transverse momentum scales we will
neglect these and concentrate on the second term, which corresponds
to the creation of a new correlated quark-gluon pair in the projectile
due to radiation from the quark.
This leads us to identify
\begin{multline}
 D_{qg}(x_q,x_g,Q^2) \approx
\frac{\as(Q^2)}{2 \pi} \ln \Lambda^2 
\ q(x_q+x_g,Q^2)
\\
\times 
\frac{1}{x_q+x_g}P_{q\to qg}\left(\frac{x_g}{x_q+x_g}\right).
\end{multline}

 With this
identification we write the contribution of \eq\nr{eq:dihadron-xs-dps4}
as
\begin{multline}
\label{eq:dihadron-xs-dps5}
\frac{\ud \sigma^{hA\to qgX}}{\ud y_q \ud y_g \ud^2 \qt  \ud^2 \kt  } 
\underset{DPS}{=} 
z(1-z)x^2 D_{qg}(zx,(1-z)x,\Lambda^2) 
\\
\times \frac{S_\perp}{(2\pi)^4}
S(\qt) \widetilde{S}(\kt).
\end{multline}
In the uncorrelated approximation where the double parton distribution
is approximated by 
\begin{equation}
D_{qg}(x_q,x_g,Q^2) \approx
x_q q(x_q,Q^2)
x_g g(x_g,Q^2),
\end{equation}
the result of \eq\nr{eq:dihadron-xs-dps5} can be identified
as the product of the single inclusive cross sections
\nr{eq:dsigmasq} and \nr{eq:dsigmasg}
as
\begin{equation}
\frac{\ud \sigma^{hA\to qgX}_{\textrm{DPS}}}{\ud y_q \ud y_g \ud^2 \qt  \ud^2 \kt  } 
= \frac{1}{S_\perp}
\frac{\ud \sigma^{hA\to qX}}{\ud y_q  \ud^2 \qt    } 
\frac{\ud \sigma^{hA\to gX}}{\ud y_g  \ud^2 \kt    } .
\end{equation}
The prefactor, which must have the dimensions of an inverse area,
is commonly denoted as $1/\seff$. In our calculation, which
assumes a dilute probe hitting a large, uniform target, this misleadingly 
named ``effective cross section'' is the same as the transverse size 
 of the target $S_\perp$. 

\section{Impact parameter profiles}\label{app:b}

Let us start (see e.g. \re\cite{Blok:2010ge}) by considering the general case of two colliding projectiles
1 and 2, of partons distributed with an impact parameter profile that we will
assume is factorized into a distribution $F_{1,2}(\bt)$ normalized as 
\begin{equation}
\int \ud^2\bt F_{1,2}(\bt)=1.
\end{equation}
We define the Fourier transform as
\begin{equation}
F(\Deltat)= \int \ud^2\bt e^{i \bt \cdot \Deltat}F(\Deltat).
\end{equation}
A single inclusive cross section is simply
\begin{multline}
\sigma^S = \int \ud^2 \bta \ud^2 \btb F_1(\bta) F_2(\btb) \sigma^{ij}
\\
=  \int \ud^2 \bt \ud^2 \bta  F_1(\bta) F_2(\bt-\bta)\sigma^{ij},
\end{multline}
where $\sigma^{ij}$ is the parton level cross section multiplied by 
the appropriate parton distribution (which we assumed factorized from the 
impact parameter profile) and $\bt$ the
impact parameter of the collision. For a double scattering the cross section can be 
written as
\begin{multline}
\sigma^D = 
 \int \ud^2 \bt \ud^2 \bta  \ud^2 \btap F_1(\bta) F_1(\btap)   
\\
F_2(\bt-\bta) F_2(\bt-\btap) 
\sigma^{ij} \sigma^{kl}
\\
= \int \frac{\ud^2 \Deltat}{(2\pi)^2} |F_1(\Deltat)|^2 |F_2(\Deltat)|^2 
\sigma^{ij} \sigma^{kl}.
\end{multline}
The term ``effective cross section'' refers to the factor
\begin{equation}
\frac{1}{\seff}= \int \frac{\ud^2 \Deltat}{(2\pi)^2} |F_1(\Deltat)|^2 |F_2(\Deltat)|^2,
\end{equation}
and has experimentally been measured as 
$\seff\approx 15\mb$ in proton-antiproton 
collisions~\cite{Abe:1997xk,Abazov:2009gc}.
The value $15\mb$ is surprisingly small (i.e. the cross section surprisingly large) 
compared to other estimates of the
proton impact parameter profile, which could lead one to question whether it includes
some additional dynamical correlations that should properly be included
in the double parton distribution itself. 
For example a Gaussian parametrization of
the TOTEM~\cite{Antchev:2011zz} $B=23.6 \gev^{-2}$ elastic cross section
$\ud \sigma/\ud t \sim e^{23.6 t/\gev^2}$ would correspond to 
$F(\Deltat) = e^{- 5.9 \Deltat^2/\gev^2}$ and $\seff \gtrsim 100\mb$. A perhaps more
relevant data point for hard small $x$ processes would be 
diffractive vector meson production
in DIS~\cite{Kowalski:2006hc,Lappi:2010dd}, where we can parametrize
$F(\Deltat) = e^{- B_D \Deltat^2/2}$ with $B_D\approx 4\gev^{-2},$ leading to 
$\seff\approx 39 \mb$. The dipole parametrization 
estimate~\cite{Frankfurt:2002ka}
for the proton two gluon form factor, 
$F_{2g}(\Deltat) = 1/(\Deltat^2/m_g^2)+1)^2$ with $m_g^2 = 1.1\gev^2$, 
similarly leads to a large value for $\seff$.

It was, however, noticed in \re\cite{Strikman:2010bg} that the experimental 
result $15\mb$ could be interpreted as the large~$x$ valence-like 
partons being  completely localized in the transverse plane. 
In this case the dilute probe form factor would be 
$F_1(\Deltat)=1$ and
\begin{equation}\label{eq:dildensb}
\frac{1}{\seff}= \int \frac{\ud^2 \Deltat}{(2\pi)^2} |F_2(\Deltat)|^2 = 
 \int \ud^2 \bt   F_2(\bt)^2 .
\end{equation}
Equation \nr{eq:dildensb}
leads to an intuitive picture of the double inclusive cross
section as an integral over the transverse plane of the target that
is directly generalizable to nuclei. For a nuclear target 
with $F_2(\bt) \sim T_A(\bt)$ this would lead to a similar impact parameter
dependence as the contributions (b) and (c) in \re\cite{Strikman:2010bg}.
 Here the target is proportional to $T_A(\bt)$ only in the 
dilute limit and assuming that the target density is $\sim T_A$ 
would be inconsistent with a nuclear modification 
factor $R_{pA}$ significantly different from one in single
inclusive scattering. However, motivated by the observation in 
\re\cite{Strikman:2010bg} that the proton-proton result of 
$\seff \approx 15\mb$ is consistent with the large~$x$ projectile being
pointlike we shall keep the interpretation~\nr{eq:dildensb}
for $\seff$. Thus the only impact parameter dependence in our
calculation is an integral over the transverse profile of 
the target, which disappears when calculating yields instead
of cross sections, as in \eqs\nr{eq:dsigmasq},\nr{eq:dsigmasg},
 \nr{eq:dAuqg-xs} and  \nr{eq:totDPS}.

\bibliography{spires}
\bibliographystyle{JHEP-2modM}

\end{document}